# Resolving discrepancies in bang-time predictions for indirect-drive ICF experiments on the NIF: Insights from the Build-A-Hohlraum Campaign

Work in progress 6/11/2025

G. F. Swadling[1a], W. A. Farmer[1], H. Chen[1], N. Aybar[1], M. S. Rubery[1], M. B. Schneider[1], D. A. Liedahl[1], N. C. Lemos[1], E. Tubman[1], J. S. Ross[1], D. E. Hinkel[1], O. L. Landen[1], M. D. Rosen[1], S. Rogers[1], K. Newman[1], D. Yanagisawa[1], N. Roskopf[1], S. Vonhof[2], L. Aghaian[2], M. Mauldin[2], B. L. Reichelt[3], J. Kunimune[3]

## AFFILIATIONS

[1]Lawrence Livermore National Laboratory, P.O. Box 808, Livermore, California 94551 USA
[2]General Atomics, San Diego, California 92121, USA
[3]Massachusetts Institute of Technology, Cambridge, Massachusetts 02139, USA
[a)]**Author to whom correspondence should be addressed: swadling1@llnl.gov**

## ABSTRACT

This study investigated discrepancies between measured and simulated x-ray drive in Indirect-Drive Inertial Confinement Fusion (ID-ICF) hohlraums at the National Ignition Facility (NIF). Despite advances in radiation-hydrodynamic simulations, a consistent "drive deficit" remains. Experimentally measured ID-ICF capsule bang-times are systematically 400-700 ps later than simulations predict. The Build-A-Hohlraum (BAH) campaign explored potential causes for this discrepancy by systematically varying hohlraum features, including laser entrance hole (LEH) windows, capsules, and gas fills. Overall, the agreement between simulated and experimental x-ray drive was found to be largely unaffected by these changes. The data allows us to exclude some hypotheses put forward to potentially explain the discrepancy. Errors in the local thermodynamic equilibrium (LTE) atomic modeling, errors in the modeling of LEH closure and errors due to a lack of plasma species mix physics in simulations are shown to be inconsistent with our measurements. Instead, the data supports the hypothesis that errors in NLTE emission modeling are a significant contributor to the discrepancy. X-ray emission in the 2 – 4 keV range is found to be approximately 30% lower than in simulations. This is accompanied by higher than predicted electron temperatures in the gold bubble region, pointing to errors in non-LTE modeling. Introducing an opacity multiplier of 0.87 on energy groups above 1.8 keV improves agreement with experimental data, reducing the bang-time discrepancy from 300 ps to 100 ps. These results underscore the need for refined NLTE opacity models to enhance the predictive power of hohlraum simulations.

## I. INTRODUCTION

Numerical modeling of indirect-drive[1,2] Inertial Confinement Fusion (ID-ICF) experiments has played a critical role in the development of experimental platforms that are now routinely capable of reaching fusion ignition at the National Ignition Facility (NIF)[3–6]. To achieve these high-performance results, experimental designs are extensively modeled using sophisticated radiation-hydrodynamic simulations[7,8]. The results of these simulations inform decisions regarding changes to key design parameters such as the hohlraum size and shape, wall material, gas fill and the details of the laser pulse shape used to drive the target.

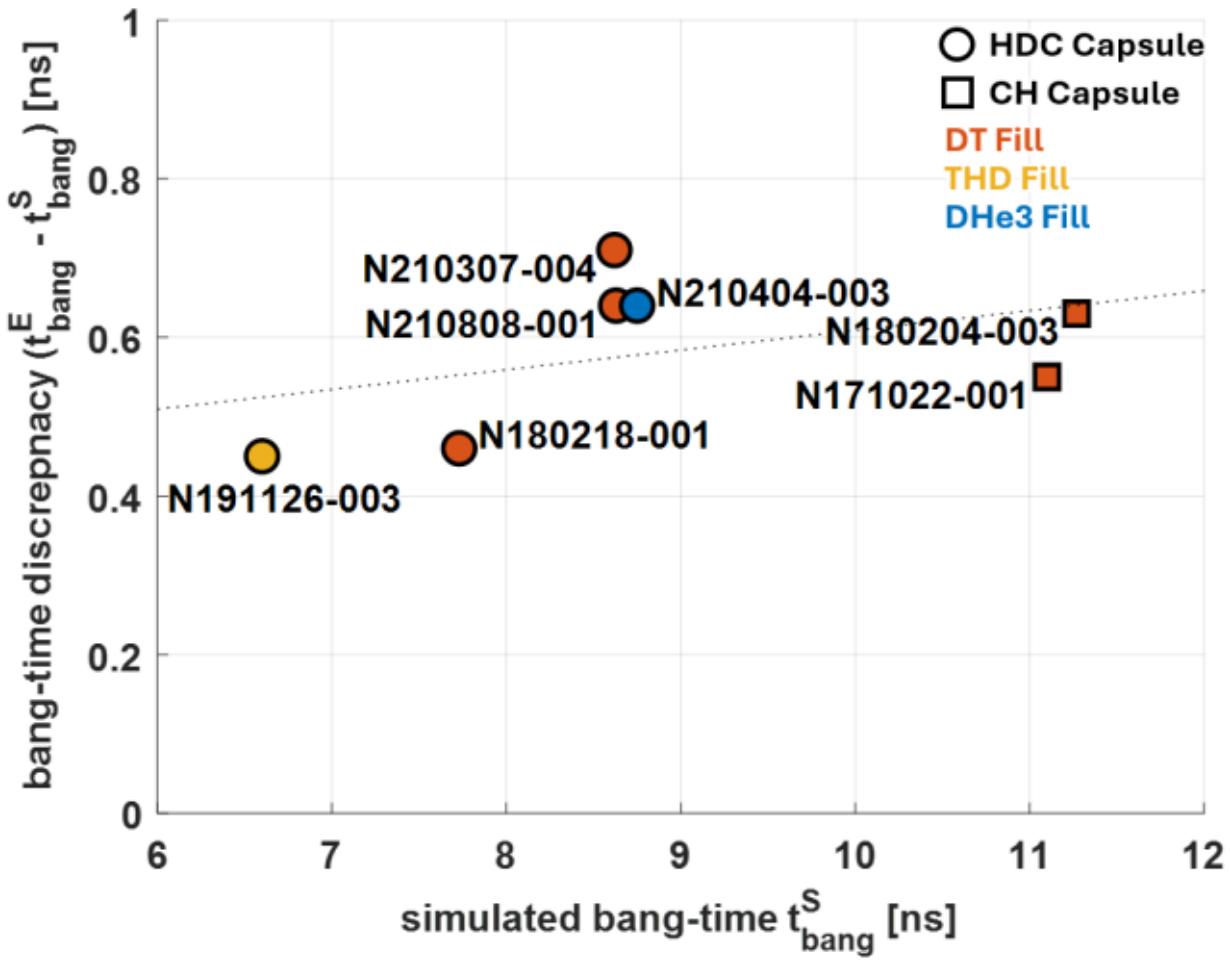

FIG. 1 Plot of the "bang-time" discrepancy between experiment and simulation for a range of ID-ICF experiments, including both CH and HDC capsules and a range of fill types. Experimentally measured bang-time is consistently ~400-700ps later than expected based on simulations. Simulations used the Lasnex Hohlraum Template discussed in section IV.

While it may be possible in principle to accurately simulate hohlraums using high-fidelity physics models in full 3D, these simulations would be extremely computationally expensive. Even with the extreme advances in supercomputer processing power that have been achieved over the past half century[9,10], it remains impractical to fully simulate all of the detailed atomic and kinetic physics that determine hohlraum performance. In practice, hohlraum simulations are routinely carried out in 2D (r,z, plane, assuming cylindrical symmetry), and critical processes such as radiation and atomic physics[11–14], heat transport[15–17] and laser plasma interactions[15,18–21] are routinely treated using reduced models. While these practical tools have been critical to achieving ignition[6], they are not yet truly predictive. Many reduced models are controlled and defined by a range of empirical parameters that must be adjusted on a platform-by-platform basis to match experimental data.

One well-known example of these tuning parameters is the use of drive multipliers[22], which reduce the laser drive power delivered to the target in simulations relative to that measured in experiments. Drive in simulations is reduced in order to match the experimentally measured "bang-time", the time of peak neutron emission from the imploded fuel capsule. Over a wide range of experiments, bang-time is consistently measured to be ~400-700 ps later than predicted by simulations (see FIG. 1)[23]. Reductions in drive as large as 15-20% are required to match experimental bang-time measurements[22], much larger than the ~2% errors in the measurement of delivered of laser energy[24]. Within the ID-ICF discipline, the requirement to use these multipliers to match experiments is commonly referred to as the "Drive Deficit" problem. Accurate predictions of the bang time are critical to estimating the implosion velocity of a given design and for correctly timing x-ray diagnostics used to measure implosion quality.

Hypotheses to explain the discrepancy between simulated and experimental bang-times include errors in the modeling of Laser Entrance Hole (LEH) closure[25–28], lack of modeling of mix between the hohlraum fill plasma & capsule ablation plasma and the hohlraum wall[29], errors or inaccuracies in modeling of the Local Thermodynamic Equilibrium (LTE) gold wall properties[30,31], the Non-LTE wall properties[13,14,22,22,32], the LEH hardware[33], undiagnostic laser coupling losses[34] and the capsule's response to the x-ray drive[35].

The continued need for empirical tuning parameters like drive multipliers in simulation codes limits how these tools can be used. While they currently provide a powerful means of informing "local" changes in the experimental parameter space, they cannot be relied upon to accurately predict the effects of large changes to experimental designs. This necessitates an iterative approach to experimental platform development which significantly increases both the cost and time required for innovation. There is therefore a strong motivation to investigate these effects with a view to improving our models and moving closer to a fully predictive hohlraum modeling capability[36].

In this paper we present the results of the Build-A-Hohlraum (BAH) campaign; a series of experiments aimed at identifying the underlying causes of the bang-time discrepancy. The experiments focused on hypotheses that might explain reduced x-ray drive in experiments as compared to simulations. The campaign consisted of a series of experiments using sub-scale targets of gradually increasing complexity. Starting from a very simple vacuum hohlraum, additional features like laser entrance hole (LEH) windows, a capsule, and a gas fill were added one by one, with the aim of revealing which of these features had the greatest effect on the x-ray drive in the hohlraum.

Overall, changes in the target design complexity had very little effect on the agreement between the experimentally measured x-ray emission of the hohlraum and simulation. These observations allow us to reject the hypothesis that mix between the hohlraum fill gas or the ablated capsule plasma and the gold walls leads to significant additional energy losses. Adding a gas fill or capsule to the target did not introduce significant additional discrepancies between the measured and simulated x-ray emission and did not result in any additional increase in the gold bubble electron temperature compared to temperatures seen in vacuum hohlraums without a capsule.

The total measured x-ray emission from our hohlraums was consistently measured 6% lower than predicted by simulations using Dante-1. More detailed analysis showed that this discrepancy was due to a 30% overprediction of emission in the higher energy 2 – 4 keV band. Spectroscopic measurements of the gold bubble region reveal a significantly higher electron temperature than predicted by simulations. Combined, these observations point to errors in the NLTE modeling of the gold bubble M-band emission. These data corroborate the conclusions of another recent paper[32,37], which reported on radiation drive measurements from "ViewFactor" targets[38]. Once this NLTE emission discrepancy is accounted for we find agreement within our measurement errors between the measured and simulated thermal LTE emission power from the hohlraum, allowing us to reject the hypothesis that errors in the LTE modeling of the gold wall are responsible for a significant portion of the drive deficit.

The data collected in the campaign indicates that simulations still do not accurately capture LEH closure. Across our suite of experiments, we see a consistent pattern where simulations overpredict the LEH size early in the drive, approximately match LEH size at peak drive and slightly underpredict size after the laser drive ends. The overpredicted LEH size ought to lead to larger LEH loses in simulations and therefore a lower hohlraum temperature and reduced coupling to the capsule. This LEH size discrepancy therefore cannot explain the historical bang-time discrepancy but may help to explain discrepancies in the temporal shape of the simulated and experimentally measured x-ray radiant intensity.

The ViewFactor study[32] concluded that agreement between experimental and simulated temperature and radiation drive measurements could be significantly improved via the introduction of an M-band opacity multiplier to artificially reduce M-band emission, effectively reducing the efficiency of laser power conversion to x-ray power ($\eta_{laser}$). Adopting a similar approach, we achieve good agreement between simulated and measured x-ray emission when using an opacity multiplier $\kappa_M = 0.87$ to reduce emissivity for hohlraum emission in energy groups >1.8 keV. Using these modified simulations to model an experiment using a standard NIF laser pulse, we found that the bang-time discrepancy could be reduced from 300 ps to 100ps, which is within the uncertainty of our measurement. We therefore conclude that the errors identified in the NLTE modeling are capable of explaining a large part of the bang-time discrepancy.

Some inconsistencies in the data remain. The total x-ray drive measured by Dante-2 was on average ~9 % higher than simulated for experiments that include the hardware associated with the LEH windows, but ~15-20 % lower compared to simulation for experiments which do not include this hardware. This discrepancy may relate to inadequacies in the modeling of the LEH hardware, but issues with the Dante-2 measurements due to partial occlusion of the line of sight by parts of the LEH hardware makes drawing strong conclusions from these measurements difficult, and further work will be required to properly resolve this discrepancy.

The remainder of the paper is structured as follows. In II we provide a more detailed discussion of the main hypotheses that have been put forward to explain the bang-time discrepancy and provide a summary of the experimental evidence from the BAH campaign both in support or against each of these hypotheses. In section III we describe the experimental setup used for the campaign and in section IV we describe the simulations used to model the BAH campaign. In section V we present more detailed discussions of the experimental measurements and comparisons with simulation. In section VI we discuss how agreement between simulation and experiment may be improved for a range of observations by introducing an opacity multiplier to reduce emission for energy groups >1.8 keV. Finally in section VII we present our conclusions.

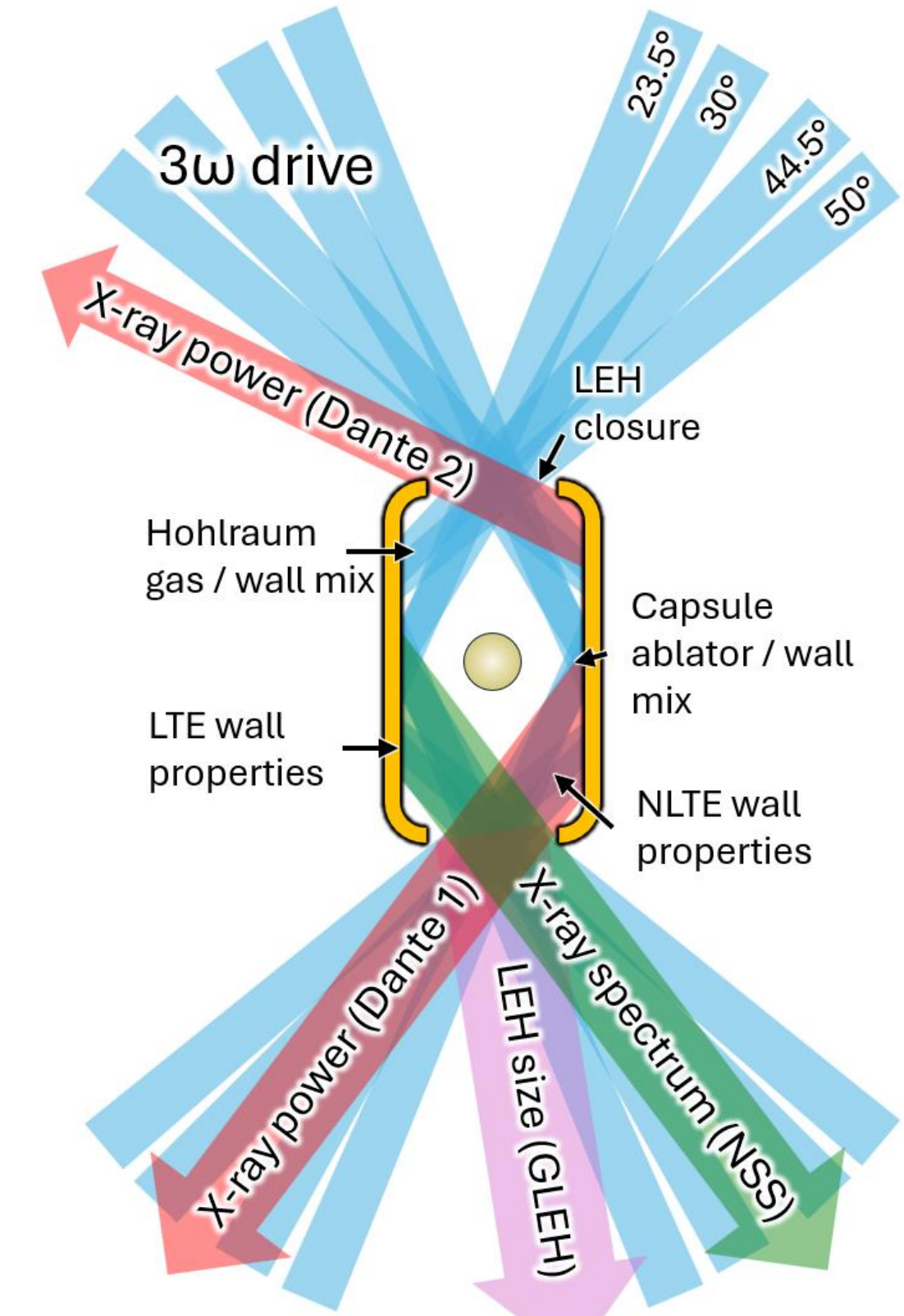

FIG. 2 Diagram of the experimental setup used in this study, illustrating the beam drive geometry and the locations of key diagnostics with respect to the hohlraum target. Targets are 5.75 mm dia. × 10.1 mm Au hohlraums with 1.8 mm dia. HDC capsules. Regions of the target associated with our various hypotheses are labeled.

## II. OVERVIEW OF HYPOTHESES THAT HAVE BEEN PROPOSED TO EXPLAIN THE "DRIVE DEFICIT"

The "drive deficit" describes the overall trend of our simulations to predict implosion times for ID-ICF capsules that are earlier than those measured in experiments. At a high level, this bang-time discrepancy could be explained by overestimates in the measurements of the laser energy delivered to the target, overestimates of the simulated laser-to-x-ray conversion efficiency, underestimates of simulated x-ray energy losses in the hohlraum, overestimates of x-ray coupling to the capsule and by errors in the modeling of the capsule response to the x-ray power delivered to it or by undiagnosed laser coupling losses. Starting from these high-level principles, the ID-ICF program has developed a broad set of more detailed hypotheses that may be able to explain some part of the discrepancy. The aim of the BAH campaign was to collect evidence to inform these hypotheses.

This section provides some physics background for each hypothesis, along with key conclusions drawn based on analysis of data from the BAH campaign. A diagram of a typical hohlraum target is provided in FIG. 2, illustrating the regions of the target where these different effects occur.

A simple power balance model of the hohlraum is helpful in informing the expected sensitivities for the hohlraum drive [2,39]:

$$\eta_{ce} f_{laser} P_{laser} = P_{cap} + P_{wall} + P_{LEH} \quad (1)$$

The laser drive power delivered to the target ($P_{laser}$) is coupled to the hohlraum at some efficiency $f_{laser}$ and converted to x-ray radiation drive at some efficiency $\eta_{ce}$ and divided between the x-ray power absorbed by the capsule ($P_{cap}$), the power absorbed by the hohlraum wall ($P_{wall}$) and the power that escapes the hohlraum LEHs ($P_{LEH}$):

$$P_{cap} = \sigma T_r^4 A_{cap} f_{cap} \quad (2)$$

$$P_{LEH} = \sigma T_r^4 A_{LEH} \quad (3)$$

$$P_{wall} = \sigma T_r^4 A_{wall} (1 - \alpha_{wall}) \quad (4)$$

Where $A_{cap}$, $A_{wall}$ and $A_{LEH}$ are the areas of the capsule, hohlraum wall and LEH respectively, and $\alpha_{wall}$ is the albedo (effectively the x-ray reflectivity) of the hohlraum wall, $\sigma$ is the Stephan-Boltzmann constant and $f_{cap}$ in the fraction of incident x-ray power absorbed by the capsule, which is simply the complement of the capsule albedo $\alpha_{cap}$.

$$f_{cap} = 1 - \alpha_{cap} \quad (5)$$

These equations can be rearranged to give an expression for the $T_{rad}$ we expect the hohlraum to reach in steady state,

$$\sigma T_r^4 = \frac{\eta_{ce} f_{laser} P_{laser}}{(A_{cap} f_{cap} + A_{LEH} + A_{wall}(1 - \alpha_{wall}))} \quad (6)$$

For the 5.75 mm diameter, 10.1 mm long hohlraums used in our experiments, typical values for these parameters are $A_{wall}$~2.2 cm$^2$, $A_{LEH}$~0.15 cm$^2$, and $A_{cap}$~0.04 cm$^2$ (at 2/3 radial convergence) $\eta_{ce}$~0.95, $f_{cap}$~0.8. We assume all delivered laser power is coupled to the target ( $f_{laser} \rightarrow 1$ ). Under this choice of units, $\sigma$ =10.26 TW cm$^{-2}$ heV$^{-4}$. An estimate for the average albedo, $\bar{\alpha}_{wall}$ that assumes constant laser drive power can be derived from the equations (4-13), (4-22), (4-23) in Lindl et al.[2],

$$\bar{\alpha}_{wall} \sim 1 - 0.44\, T_0 \tau^{-0.46} \; [heV, ns] \quad (7)$$

, where $T_0$ is the radiation temperature at 1 ns in the hohlraum and $\tau$ is the overall drive duration. This estimate is based on an analytical model of the propagation of the non-linear heat wave into the hohlraum wall. Drive temperature at 1 ns can be approximated as 20 eV less than the peak for our experiments, based on a survey of our time resolved measurements. Solving equations (6) & (7) at $P_{laser}$ = 400 TW, we expect the hohlraum to reach $T_{rad}$~288 eV, with an $\alpha_{wall}$ ~ 0.84, $P_{LEH}$ ~ 108 TW, $P_{wall}$ ~ 250 TW and $P_{cap}$~23 TW. This calculation illustrates how important it is to properly capture the various hohlraum losses. LEH and hohlraum wall losses are both significantly larger than the power coupled to the capsule.

### *1. LTE wall properties – thermal hohlraum wall emission*

Energy losses to the hohlraum wall are parameterized by the albedo ($\alpha_{wall}$), or average x-ray reflectivity of the wall. The albedo is a time dependent parameter, sensitive to the propagation speed of the radiation-driven thermal wave that diffuses into the hohlraum wall[30,31]. The velocity of this "Marshak wave"[40] is sensitive both to the drive power delivered to the hohlraum wall, the opacity of the gold and the specific energy of the wall material[31]. The opacity in the wall is modeled using an LTE atomic physics model while the specific energy depends on the accuracy of the LTE equation of state for gold, itself sensitive to the details of ionization state in the warm dense hohlraum wall material.

Errors in the modeling of either the opacity or specific energy could lead simulations to overestimate the hohlraum wall albedo, leading to an overestimate of the hohlraum radiation temperature and therefore an overestimate of the power coupled to the capsule. By substituting equation (6) into equation (2) and taking the derivative with respect to wall albedo $\alpha_{wall}$ we can estimate the fractional change in power absorbed by the capsule as a function of $\alpha_{wall}$.

$$\frac{1}{P_{cap}} \frac{dP_{cap}}{d\alpha_{wall}} = \left[1 - \alpha_{wall} + \frac{A_{cap}}{A_{wall}} f_{cap} + \frac{A_{LEH}}{A_{wall}}\right]^{-1} \quad (8)$$

For the drive parameters discussed above, $\frac{1}{P_{cap}} \frac{dP_{cap}}{d\alpha_{wall}}$ ~ 4.5, illustrating how sensitive power coupled to the capsule is to $\alpha_{wall}$; a ~0.02 absolute overestimate of $\alpha_{wall}$ is all that is required to explain a 10% overestimate of x-ray power coupled to the capsule in our simulations.

To address this hypothesis, we carried out a campaign-wide survey of x-ray emission from a range of different hohlraum designs. Errors in the LTE physics should be revealed through discrepancies between the measured and simulated spectral intensities of the thermal portion (defined as photon energies < 1.8 keV) of the x-ray emission[41] measured by Dante. Overall, the BAH campaign does not find strong evidence for these discrepancies. While the total x-ray drive was found to be on average ~6% low across the campaign, after accounting for discrepancies in the higher-energy M-band emission we find excellent agreement between simulation and experiment for the total thermal x-ray radiant intensity. In conclusion, the data from this campaign does not show clear evidence of LTE modeling errors. More details and discussion of this analysis are presented in section V.B.

### *2. Non-LTE wall properties – non-thermal hohlraum wall emission*

While the majority of laser drive energy is converted to x-rays in the high-density plasma close to the gold wall, a significant portion of the total x-ray power originates as non-thermal emission from the higher temperature, lower density gold plasma that expands out from the hohlraum wall. This expansion is fastest at the points on the wall where the heater beams hit, leading to the formation of a region commonly referred to as the "gold bubble". The lower density and higher temperature (~3 keV) of this plasma compared to the conditions at the hohlraum wall (~300 eV), mean that it can no longer be accurately treated using models that assume local thermodynamic equilibrium. Instead these regions must be modeled using an NLTE atomic physics model[13,14].

The potential for NLTE models to impact simulated hohlraum performance were highlighted by early NIF experiments which revealed deficiencies in the XSN[11] average atom atomic physics model used at that time. Simulations using the XSN model and a restrictive 0.05 heat flux limiter resulted in 20-30% underestimates of hohlraum x-ray emissivity[42,43]. These simulations overestimated the specific energy stored in the coronal Au wall plasma, reducing the effective x-ray conversion efficiency. Adoption of a Detailed Configuration Accounting (DCA) model[13] and a significantly less restrictive heat flux limiter brought agreement between simulated and experimentally measured total x-ray emission to within 5%.

Residual errors in the details of the NLTE modeling of the gold bubble plasma opacity could result in changes in the power and spectral content of the overall x-ray emission from the hohlraum. This region is predominantly responsible for the generation of non-thermal gold "M-band" (E > 1.8 keV) component of the hohlraum x-ray drive. The complex collisional-radiative physics required for accurate NLTE modeling of high-Z plasmas[13] are computationally demanding. Generally, NLTE models are believed to be less accurate[44] than the simpler LTE models used to calculate the thermal portion of the hohlraum wall emission.

Measurements using the new Dante-1 2 - 4 keV multilayer mirror channel diagnostic[45] and gold bubble temperature measurements using the NIF Survey Spectrometer (NSS)[46] were used to investigate this hypothesis. Measurements of the 2 - 4 keV emission were found to be consistently 30% lower than predicted by simulations. This discrepancy corresponds to an overprediction of ~5% total hohlraum x-ray emission. The lower than predicted m-band emission is accompanied by gold-bubble temperatures that are consistently ~31 % higher than predicted. Together, these data point to errors in the NLTE modeling which are leading to overestimates of the gold bubble emissivity.

These results corroborate conclusions from a recently published study of ViewFactor hohlraums[32]. In that paper an opacity-multiplier of $\kappa_M = 0.8$ was used to improve the agreement between simulations and experiments. This multiplier reduces the plasma opacity for photon energies > 1.8 keV, effectively reducing $\eta_{ce}$. For the majority of the Build-A-Hohlraum experiments we found that an opacity multiplier of $\kappa_M = 0.87$ brought the simulated x-ray emission into good agreement with experiment. For experiment N230207-1, which was the only BAH experiment to use a realistic ignition-type laser pulse, the $\kappa_m = 0.87$ multiplier reduced the bang-time discrepancy from 300 ps to 100 ps, while closely matching the experimentally measured x-ray emission and gold bubble plasma conditions.

As was seen in the ViewFactor experiments[32], our experiments provide some evidence that the best choice of $\kappa_m$ varies with the LEH size. The opacity multiplier as currently implemented is a blunt instrument, reducing emission for radiation > 1.8 keV from all sources in the hohlraum, including thermal contributions from the dense wall plasma. This apparent sensitivity to LEH size therefore likely reflects changes in the fraction of the Dante line of sight that views the hot gold plasma in the gold bubble. Further improvements in NLTE opacity modeling will be required to develop a reliable and consistent NLTE emission model that accurately matches M-band emission for all targets.

### *3. Laser Entrance Hole (LEH) closure*

In ID-ICF experiments, laser energy is delivered to the target through the Laser Entrance Holes (LEH) at either end of the hohlraum, as illustrated in FIG. 2. The lip of the LEH is heated during the experiment, both directly by the low-intensity halo of the drive laser beams and indirectly via x-ray emission from the hohlraum walls. This heated material expands into the LEH, reducing its effective size. In experiments with an LEH membrane and gas fill, the expansion of the LEH lip is tamped by the pressure due to the direct laser heating of the plasma generated from this material, complicating accurate modeling.

Closure of the LEH has the potential both to increase x-ray coupling to the capsule via reduced LEH losses (see equation (3)) and to decrease laser coupling to the target by blocking the propagation of the laser beams into the hohlraum. The size of the LEH also sets the area of the hohlraum wall viewed by external diagnostics such as Dante. Discrepancies in LEH size between experiment and simulation therefore have the capacity to impact comparisons of the x-ray drive. For this reason LEH closure has been studied intensively for a number of years[25–28]. Discrepancies in the LEH closure rate have been identified in previous studies[26], but these deficiencies were thought to have been addressed via changes to the heat transport model used in ID-ICF simulations[27].

The BAH campaign used the Gated LEH (GLEH)[27,28] x-ray imager on the Static X-ray Imager (SXI)[47,25] diagnostic line of sight to measure the time-resolved diameter of the LEH in a subset of experiments. Overall, the data shows that LEH closure was faster in experiments than predicted in simulations. This difference in closure rate appears limited to a perhaps ~100 µm difference in diameter at peak radiative intensity. Again, substituting equation (6) into equation (2) and differentiating, we can derive the sensitivity of capsule drive to this discrepancy.

$$\frac{1}{P_{cap}}\frac{dP_{cap}}{dA_{LEH}} = -\left[A_{cap}f_{cap} + A_{LEH} + A_{wall}(1-\alpha_{wall})\right]^{-1} \tag{9}$$

For our target parameters $\frac{1}{P_{cap}}\frac{dP_{cap}}{dA_{LEH}} \sim -2\ \mathrm{cm}^{-2}$. At a nominal radius of 3mm a 100 µm discrepancy reflects a $\sim 10^{-2}\ \mathrm{cm}^{-2}$ difference in area and therefore could drive a ~2% reduction in power coupled to the capsule in our simulations. The overprediction of LEH size therefore cannot explain the bang-time discrepancy

The effect on power loss from the LEH can be derived similarly using equation (3),

$$\frac{1}{P_{LEH}}\frac{dP_{LEH}}{dA_{LEH}} = -\left[A_{cap}f_{cap} + A_{LEH} + A_{wall}(1-\alpha_{wall})\right]^{-1} + A_{LEH}^{-1} \tag{10}$$

where the first term reflects the reduced hohlraum temperature and the second term reflects in the increase in the LEH size. Using our hohlraum parameters gives $\frac{1}{P_{LEH}}\frac{dP_{LEH}}{dA_{LEH}} = -2 + 6.6 = 4.6\ \mathrm{cm}^{-2}$, and therefore an expected increase in simulated x-ray emission of ~4%.

Agreement in the LEH size at the time of peak emission is reasonably good, and therefore is unlikely to have a big impact on the comparisons of peak emissivity presented in this paper. Details of this analysis are provided in section V.C.

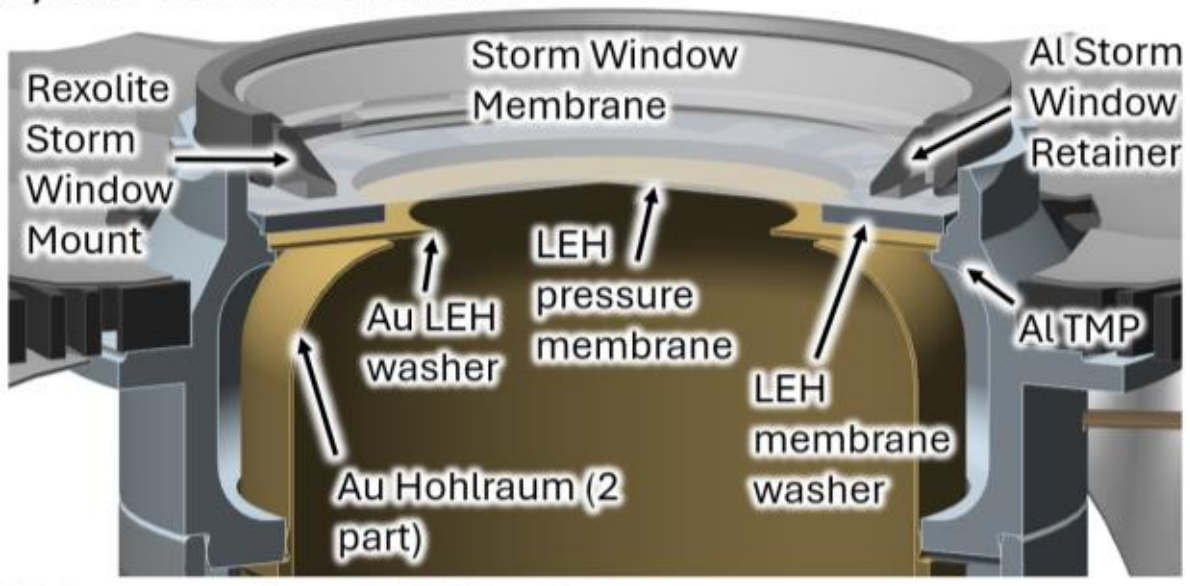

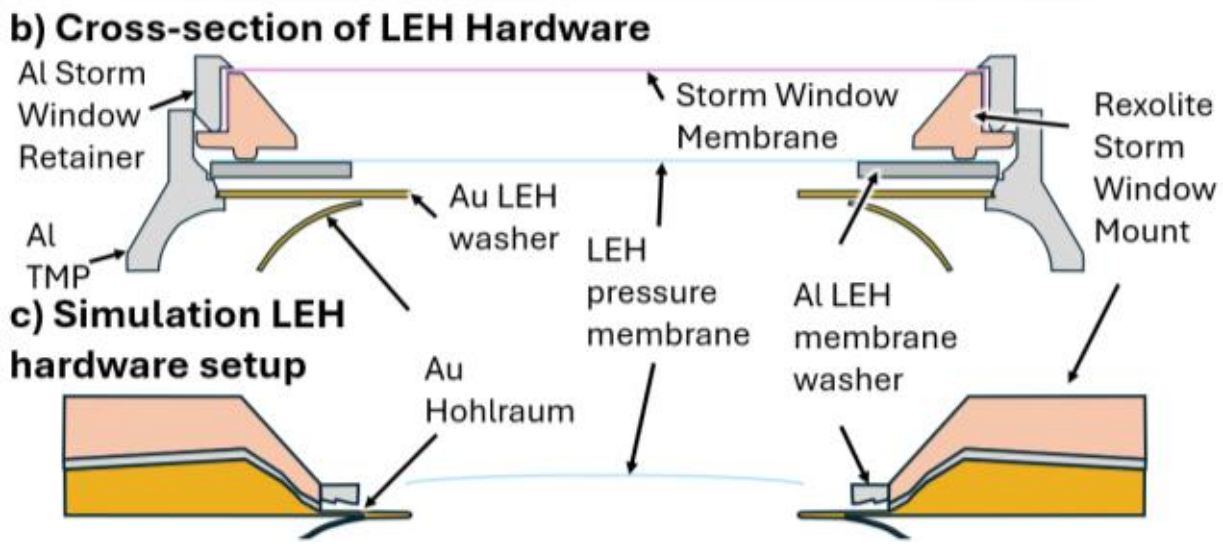

FIG. 3 Detailed view of the LEH hardware set used on more complex targets. a) Cut away cad model showing LEH hardware. b) Cross section of the LEH hardware showing the layout and materials. The LEH pressure window is shown unloaded – when fielded this window will bow outwards due to the pressure of the He hohlraum fill. c) Region definitions used in Lasnex simulations. This captures the main features of the LEH hardware. The storm window membrane and Al strom window retainer are not included.

## 4. Modeling of the LEH membrane and associated hardware

ID-ICF targets are typically fielded with a hohlraum gas fill. There are three main reasons for this. Firstly, the plasma formed via the ionization and heating of this gas tamps the expansion of the laser-heated gold walls, extending the time available to deliver laser energy to the hohlraum wall and maintaining control of low-mode implosion shape. Secondly, the gas fill provides a medium for laser-plasma interactions, allowing controlled crossbeam energy transfer (CBET) to be used to transfer energy from outer to inner beam cones and control implosion symmetry without the need to increase the LEH size to allow for cone repointing[3,48,49]. Thirdly, this gas provides a medium to facilitate thermal conduction between the capsule and the hohlraum, which is necessary if we want to field cryogenically cooled nuclear-fueled capsules. To contain this gas prior to the start of laser heating, the hohlraum LEHs are sealed using a thin membrane. The standard membrane used on high-performance ID-ICF targets is a 500 nm thick polyimide film flash-coated with 25 nm of aluminum to shield the capsule from infrared radiation that leaks into the cryogenic shroud via the alignment window.

LEH membranes are susceptible to the growth of ~1 μm scale layers of residual target chamber gases that condense on their surface due to the extremely low temperatures at which they are fielded[50]. These condensates can perturb delivery of laser energy to the target. To mitigate this, a second protective "storm window" membrane is installed ~700μm outside the pressure membrane[51], preventing residual gases from reaching the LEH membrane. The membrane is thermally insulated from the target, allowing it to be warmed by infra-red (IR) radiation to a temperature where condensates do not form. It consists of a 100 nm thick polyimide film with a 32 nm carbon coating to enhance IR coupling to the film.

On integrated ID-ICF targets, the assembly consisting of the LEH insert (a gold washer), the LEH pressure membrane glued to an aluminum mounting washer and the storm window assembly constitute the "LEH hardware". As can be seen from the CAD images and cross-sections provided in FIG. 3, the combined set of hardware is complex. Accurate modeling is essential to enable useful comparisons to ID-ICF experiments. The expansion and interaction of the plasma generated during blowdown of these membranes set the initial plasma conditions for the growth of laser-plasma interaction effects such as CBET, backscatter and pondermotive filamentation that can have critical effects on target performance. As an example, CBET from outer to inner beams in the plasma formed via the ablation of the LEH membranes was eventually identified as the cause of unexpectedly prolate implosions in experiments studying near-vacuum hohlraums[33]. This effect could not be predicted without including the LEH hardware in the Lasnex simulations.

FIG. 3 c) shows the model for the LEH hardware set currently used in Lasnex simulations. While it captures many features of the hardware it does not include the aluminum retaining ring or the storm window membrane. The lack of the retaining ring proved to be important when it came to comparing simulated and experimental signals for the Dante-2 x-ray spectrometer, as this ring blocks ~27% of x-ray emission from the LEH along the Dante-2 line of sight. This is discussed in greater detail in V.B.

Experiments were conducted both with and without the LEH hardware. Addition of the LEH hardware had no discernable effect on the agreement between Dante-1 radiative intensity measurements and simulation, however significant changes were seen in the agreement between Dante 2 measurements and simulation, even after corrections for occlusion by the Al storm window retaining ring. For targets without LEH hardware, the peak x-ray radiative intensity measured by Dante 2 was systematically 15-20% lower than simulation, while for experiments with the LEH hardware set the experimental measured x-ray emission was 5 – 10 % higher than simulation after correction. This effect is not currently understood. Since the Dante-1 diagnostic has a better line of sight into the hohlraum than Dante-2, we have leant more towards the Dante-1 data in developing our conclusions. More details are provided in section V.B, where there is further discussion of this discrepancy.

## 5. Mixing between the hohlraum wall and Hohlraum gas fill / ablated capsule plasma

As discussed in the previous section, gas fills (He) are fielded in ID-ICF hohlraums for a range of reasons. Hohlraum gas fill densities have varied significantly over the life of the ID-ICF program, from a ~1 mg cm$^{-3}$ gas fill[52,53] used in the initial National Ignition Campaign, which support experiments using up to a 20ns drive pulse, up to maximum of 1.6 mg cm$^{-3}$ in the following "high foot" campaign[54]. These "high gas fill" targets were plagued by unpredictable LPI effects such as cross-beam energy transfer and backscatter; after focused study[55] the ~0.3 mg cm$^{-3}$ "low gas fill" was adopted for current ignition target designs[3,5]. This choice is a compromise designed to keep LPI predicable while maintaining control of late time low-mode implosion symmetry as the hohlraum wall starts to interfere with inner beam propagation[56–58].

The ablation and expansion of the laser heated gold wall drives a piston of Au plasma into the hohlraum gas fill plasma and mixing of this Au with the He plasma is expected to occur at the interface between the two materials. This type of species mixing physics is not typically captured in radiation-hydrodynamics codes[29]. It has been shown theoretically that mixing of the high-Z hohlraum wall plasma with the low-Z hohlraum gas fill plasmas could lead to enhanced electron heating in the mix layer[29]. This enhanced temperature would come at the expense of x-ray conversion efficiency, with an upper limit estimated at around 100 kJ of lost drive energy. A similar effect could also be driven by mixing of ablated capsule plasma with the wall, particularly at later times in the experiments where this plasma has largely displaced the initial hohlraum fill plasma near the hohlraum waist, or in hohlraums that operate with a near-vacuum hohlraum fill[33]. For an experiment with a 2 MJ drive, a 100 kJ energy

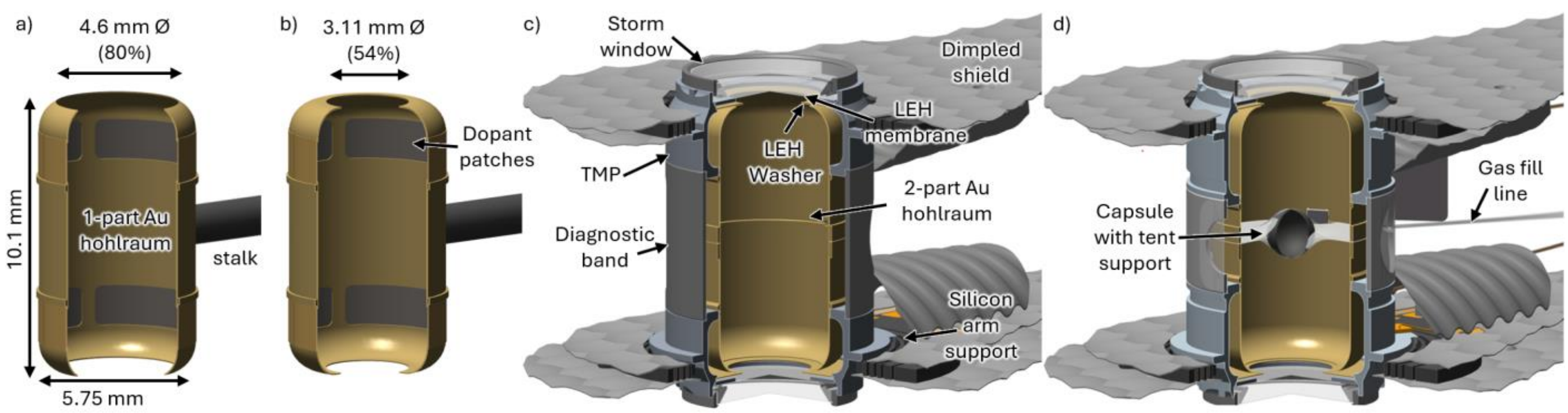

FIG. 4 Targets fielded in the Build-A-Hohlraum campaign varied significantly in complexity, from simple monolithic vacuum hohlraums designs such as a) and b) to more complex designs such as c) and d), which more closely match the state of the art target designs for current high yield experiments on the NIF.

loss due to mix would contribute ~5% out of the ~15-20% total losses we are seeking to explain.

The mix hypothesis was addressed in the BAH campaign by comparing x-ray emission measurements from targets both with and without capsules and hohlraum gas fills. Over the course of the campaign no significant changes in the agreement between experiment and simulation were observed on the addition of either a gas-fill or a capsule. While spectroscopic measurements show that the electron temperature in the gold bubble is higher than predicted by simulations, this enhanced temperature is also seen in vacuum hohlraum targets where there should be no mix effect. Our data therefore does not support the mix hypothesis. We conclude that unmodeled mix effects are not a leading contributor to the Drive Deficit.

#### *6. Capsule hydrodynamic modeling*

The details of capsule implosion modeling may also have a significant impact on the simulated bang-time. Examples of potential sensitives include the equation of state[35] for the capsule ablator and the details of x-ray energy deposition physics in the doped ablator. Under this hypothesis, it is proposed that it is not that the simulation overpredicts the x-ray drive generated by the hohlraum, but instead that the coupling of that drive to the capsule and the effect of that drive on the capsule is incorrectly modeled.

This hypothesis is not directly addressed by the BAH campaign, but we have made several measurements of bang-time where possible to allow comparison between observed x-ray drive discrepancies and actual bang-time discrepancies.

#### *7. Undiagnosed laser coupling losses*

When discussing the bang time discrepancy, we typically refer to the discrepancy that remains after accounting for the as-delivered laser pulse and any measured coupling losses. Nevertheless, systematic undiagnosed coupling losses or errors in laser power/ energy measurements still have the potential to contribute to the bang-time discrepancy. This fact is highlighted by the routine use of laser drive multipliers to reduce delivered laser energy in simulations to correctly predict bang times for experimental design purposes.

Laser delivery to the target is measured by a range of laser power and energy diagnostics and is believed to be accurate to within 2%[59], and coupling to the target is diagnosed using a range of backscatter diagnostics[60]. Coupling losses can be caused by both backscatter from LPI processes and from glint – specular reflection of laser light from the hohlraum wall at early times. Measurements of LPI backscatter using the existing diagnostic suite lead to a coupling uncertainties of ~±2%, and recent experiments[34] using witness beams have demonstrated that coupling losses from glint are <2%. These uncertainties have been shown to be too small to explain the majority of the bang-time discrepancy.

The BAH campaign did not directly investigate coupling losses. Coupling to targets was measured throughout the campaign, and fell in the range 0.96 – 0.99. Simulations used as comparisons for each experiment use the as-shot laser power and coupling losses in order to remove laser coupling effects from comparisons.

## III. EXPERIMENTAL SETUP

### A. Variations in target design

FIG. 2 provides a basic schematic of the experimental target and diagnostic configuration used throughout the campaign. All the targets in this campaign used 5.75 mm inner diameter Au hohlraums with an internal length of 10.1 mm. These sub-scale hohlraums have ~0.9 × the linear dimensions and 0.73 × the volume of those used in current high yield experiments[3], allowing experiments to reach similar radiation temperatures using significantly reduced laser power and energy. Lines of sight to key diagnostics are illustrated in FIG. 2; x-ray radiant intensity is measured using Dante-1 and Dante-2, gold bubble plasma conditions are probed with the NSS spectrometer and LEH size is measured using GLEH. These diagnostics are discussed in greater detail in section V.

Key variations in the target design are illustrated in FIG. 4, ordered from the simplest on the left to the most complex on the right. The specific variations in these design elements are tabulated on an experiment-by-experiment basis in Table 2.

The simplest target, FIG. 4 a), uses a one-part hohlraum design, mounted in the target chamber on a stalk. This target has no gas-fill, no LEH membranes and has a large LEH (4.6 mm, 80% of the hohlraum diameter). Manganese and iron dopant patches (20% fraction) were installed on these targets to facilitate dopant spectroscopy measurements of the hohlraum plasma conditions.

For the next target design, FIG. 4 b) the size of the LEH was reduced to more closely match targets used in ID-ICF experiments while maintaining the otherwise simple design (3.1 mm, 54% of the hohlraum diameter). This target was fielded both with and without dopant patches to test the effect of these patches on the x-ray conversion efficiency – the dopant patches were found to have no measurable effect on x-ray emission measurements from the target.

The third target design, FIG. 4 c), switched to the more complex design architecture used for most ID-ICF targets[61]. The hohlraum is now a two-part design, consisting of separate upper and lower halves, mounted in an aluminum thermo-mechanical package (TMP) and surrounded by an aluminum diagnostic band. These targets include the full LEH hardware set; a more detailed view of this is provided in FIG. 3. The LEH diameter (3.1 mm) is now defined by a separate gold washer, and both the LEH pressure membrane and the storm window assembly are installed. The LEH pressure membrane is a 500 nm thick polyimide film overcoated with 25 nm of Aluminum. The storm window is a 100 μm polyimide film with a 32 nm overcoat of carbon. A single experiment was conducted with the storm window assembly removed (Rexolite mount, membrane and aluminum retaining ring) to test the effect of the hardware on the Dante-2 measurement. These more complex targets are mounted in the chamber using a pair of silicon thermal arms to facilitate cryogenic cooling. Dimpled shields cover exposed flat surfaces to scatter any specular reflections of unconverted 1ω light that might otherwise back-propagate into a beam port and damage the laser. This target design was fielded without a hohlraum gas fill.

Finally, the target design shown in FIG. 4 d) adds a capsule, mounted in the center of the hohlraum on a tent. This design was fielded both with and without a He hohlraum gas fill, and with and without a capsule gas fill. Gas filled targets were fielded at cryogenic temperatures.

### B. Variations in laser drive used in the campaign

Targets were driven using the NIF high energy laser[62,59], delivering ~ 900 kJ of laser energy to the target using 192 beams organized into 48 quads arranged into 8 cones split between the 2 hemispheres at angles 23.5°& 30° ("Inners", 8 quads per side) and 44.5° & 50° ("Outers", 16 quads per side) from the hohlraum axis. The laser pointing used to drive the targets is summarized in Table 1.

| Beam Cone | Beam pointing in quad coordinates [mm] | | |
|---|---|---|---|
| | Quad r | Quad z | Beam Defocus |
| **50°** | 0 | ±5.01 | 0 |
| **44.5°** | 0 | ±5.20 | 0 |
| **30°** | 0.6 | ±3.63 | 0 |
| **23.5°** | 1.09 | ±3.09 | 0 |

Table 1 Laser pointing used for this experimental campaign, specified in terms of the radius (r) and height (z) each ring of quads is pointed to in a cylindrical coordinate system in which the origin is the hohlraum center and the z axis is the hohlraum axis. The quad pointing azimuthal angle $\phi$ is defined as the back-azimuth of the quad port location azimuth. Upper beams point to the positive z locations and lower beams point to the negative z locations.

The laser pulse shapes used to drive targets in the BAH campaign are plotted in FIG. 5. All but one of the experiments were driven using ~ 2 ns duration square laser pulses. These pulses had a maximum power of 400 TW and total energy of ~900 kJ. Vacuum hohlraums were driven by the pulse shape shown in a). In experiments that included an LEH membrane a "toe" and "foot" were added to the inner beam laser pulse to precondition the LEH plasma before the delivery of the main laser pulse[63], as illustrated in b). Finally, one experiment was conducted using a multi-tiered three-shock laser pulse design plotted in c). This pulse more closely resembles those used in contemporary high yield ID-ICF experiments.

## IV. MODELING OF EXPERIMENTS

Simulations of this suite of experiments were performed using the radiation hydrodynamic code Lasnex[7]. The code was configured using the Lasnex Hohlraum Template (LHT)[64], a "best-effort", version-controlled common model used by the LLNL ID-ICF program. The LHT remains under active development with continuous updates to recommended settings based on experimentally informed best physics practices balanced by practical computational expense. This common model exists to provide a reference setup allowing for direct comparison of simulations run by the many different scientists within the ID-ICF program. Previous mesh resolution studies have demonstrated that simulations using the LHT framework are fully converged is space, time and energy[36].The aim of the BAH experiments is to generate data that can be used to propose future improvements to this common model, with the aim of improving predictive capability.

For the simulations reported here, heat transport is simulated using the flux-limited Spitzer–Harm model with a flux limiter of $f = 0.15$. This choice was motivated by optical Thomson scattering measurements of laser irradiated gold spheres[14,16,17]. A magnetohydrodynamic (MHD) description[65] that also includes heat restriction due to self-generated fields can also be employed and will be mentioned explicitly in simulations where it is adopted. Recent experimental work shows that combining an additional opacity multiplier for higher energy (>1.8 keV) photons with MHD matches plasma condition measurements within the gold bubble and x-ray measurements[32]. Here, these model variations are explored and will be mentioned explicitly when used. When using MHD, a multiplier on the Nernst term can be varied. Two choices are made when employing the MHD model here. The first sets this multiplier to 0.1. This choice is made based on previous work[66]. The second simply sets the Nernst multiplier to unity, adopting a purely local MHD description. Since the MHD model has little impact on drive[65], variations in these choices are discussed specifically in the context of comparisons to the spectroscopic measurements later in the manuscript.

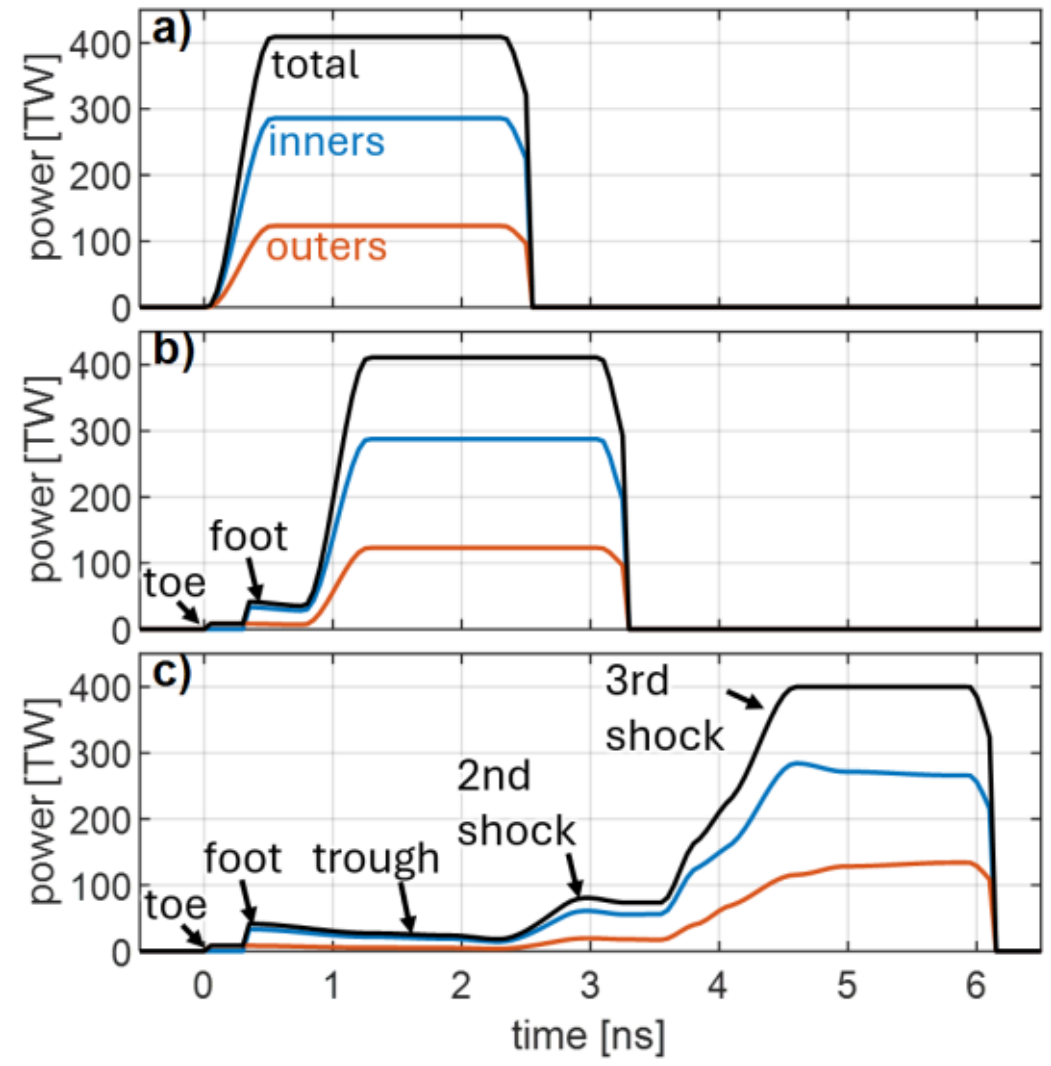

FIG. 5 Drive laser pulses used in this experimental campaign. a) Pulse used for vacuum hohlraum targets. b) Pulse used for targets with an LEH membrane. A small toe and foot are added to precondition the LEH membrane plasma prior to the start of the main drive. c) A more realistic tiered three-shock laser pulse was used in one experiment to test sensitivity of the measured x-ray drive to the more complex, shaped pulse.

An LEH hardware model is included in the simulations when appropriate. This model includes the LEH membrane and washer and adequately captures the features near the LEH (see FIG. 3). As the meshing moves away from the LEH, the engineering features are only approximately captured. This is for practical considerations in performing the simulations, and the approach could be improved to better describe the LEH hardware. The storm window itself is not included nor is the associated aluminum storm window retaining ring.

An LTE opacity model is used in regions where the plasma temperature $T_e < 300$ eV, while an NLTE model is used in hotter regions[13,14]. The NLTE model has been under continuous development with various advances over the years[36,67]. For this study, the inline models as of 2020 are used. The cross-beam-energy transfer is also modeled with a $\delta n_e/n_e$ saturation clamp[48] of 1%. Simulations used the as-shot laser pulses measured in each experiment and the as-shot target metrology to specify the initial conditions.

## V. EXPERIMENTS RESULTS

This section presents a more detailed discussion of the key experimental results throughout the campaign. Table 2 provides a summary of key experimental and simulation metrics.

### A. Laser coupling to the targets

The total amount of laser power and energy absorbed by the hohlraum target is determined based on measurements of the laser energy that is delivered to the target by the laser system and measurements of the laser energy that is scattered out by the target via processes such as SBS, SRS and glint. The simulations discussed in this paper use the experimentally measured, as-delivered laser pulse to drive the target and apply the measured glint and backscatter to reflect the measured coupling of that laser power to the target. Coupling losses are therefore not typically considered to form a part of the "drive deficit", however it should be noted that significant systematic errors in these measurements could still be a contributor to an apparent x-ray drive deficit.

The laser drive delivered to the target is measured using the suite of NIF energy and power diagnostics[24]. The energy delivered to our targets is tabulated in Table 2. Uncertainties in the total measured energy are ~2%[24]. Laser backscatter and glint from the targets were diagnosed using the full aperture backscatter (FABS) and near-backscatter imager (NBI)[60] diagnostics. Overall, laser coupling to the targets was very high throughout the campaign, with most experiments measuring coupling fractions of 0.98 - 0.99. The lowest measured coupling fraction was 0.96. The majority of the uncoupled light was "glint"- laser light reflected from the hohlraum wall before it expands and reaches full absorbance[2,68]. This is concluded based on analysis of the spectrum of the light recorded by the FABS SBS spectrometer, which consistently showed only an early time, blue shifted feature. This blue shift is consistent with the Doppler shift expected from a glinting reflection of the drive beam from the expanding Au wall of the hohlraum[69]. For SBS backscatter the signal seen by FABS ought to be red shifted[70]. Since glint is expected to be quite diffuse the total glint had to be estimated from the FABS SBS signal based on ray traced modeling of a reflection off the cylindrical hohlraum wall. Using this method, total glint was estimated to be ~22x of the measured glinted energy. Even with this large multiplier the estimated glint losses were small compared to the drive and insufficient to explain a significant portion of the Drive Deficit.

While simulated glint in the standard simulation framework is known to underpredict glint, the discrepancy should be energetically small[23].

## B. X-ray power measurements

The x-ray power emitted from the hohlraum is measured using the "Dante" time-resolved, multi-channel, filtered-diode soft-x-ray spectrometer / bolometers[71,72,45,73]. The NIF has two Dante diagnostic lines of sight, as illustrated in FIG. 2. The Dante-1 line of sight is at an angle of 36.75° from the hohlraum axis while Dante-2 is at 64°. These diagnostics measure both the time-resolved total x-ray radiant intensity $I[\mathrm{W\,sr^{-1}}]$ and spectral shape of the soft x-ray emission from the target along their respective lines of sight, allowing for detailed comparisons with simulations. Examples of these data are provided in FIG. 6. FIG. 7 a) shows the interior of a hohlraum as viewed along these two lines of sight. The Dante-1 view includes a portion of the wall directly heated by the outer laser beam ring, and a portion of the wall heated by the inner beams. As the experiment proceeds the gold bubble expands out to almost fill the Dante-1 line of sight. The Dante-2 view mainly sees a portion of the hohlraum wall that is not directly heated by the laser beams. The edge of the outer beam ring is just visible at the edge of the LEH.

### 1. *Key metrics for the measured x-ray drive*

The raw Dante data are time-resolved voltage traces measured by the arrays of x-rays diodes. X-ray filters and grazing incidence mirrors are used to control the spectral content of radiation delivered to each diode, providing energy-resolved coverage from 50eV to 20keV. The traces are calibrated and fitted with a radiation model to produce a time-dependent inference of the source spectral intensity $dI/dE\ [\mathrm{W\,sr^{-1}eV^{-1}}]$[73], as illustrated in FIG. 6. The data is then further distilled into a set of key metrics for the x-ray drive, which can be used to identify trends across the dataset. A subset of these metrics is tabulated in Table 2. The radiant intensity measured by Dante is calculated:

$$I = \int_0^{13keV} \frac{dI}{dE}\,dE\ \ [\mathrm{W\,sr^{-1}}] \qquad (11)$$

The uncertainty in the absolute experimental determination of $I$ is 5 % for Dante-1 and 10 % for Dante-2. Radiant intensity can be converted to a drive temperature $T_r$ to facilitate normalized comparisons of the drive between targets with different LEH sizes:

$$T_r[eV] = \frac{k_B}{e}\left(\frac{4I_{e\Omega}}{\sigma \emptyset_0^2 \cos(\theta_{view})}\right)^{0.25}\ [\mathrm{SI}] \qquad (12)$$

Where $\emptyset_0$ is the initial LEH diameter, $\theta_{view}$ is the viewing angle for the Dante diagnostic and $\sigma$ is the Stephan-Boltzmann constant. The initial LEH diameter is used to convert both experimental and simulated radiant intensities to drive temperatures throughout the experiment to maintain a fixed basis for comparison. Experimental uncertainties in $T_r$ simply propagated from uncertainties in $I$; 1.25% for Dante-1 and 2.5 % for Dante-2. Table 2 tabulates the peak values of these parameters for Dante-1 and Dante-2. Superscripts E & S are adopted to indicate experimental and simulated values and subscripts 1 & 2 for the Dante-1 and Dante-2 lines of sight respectively.

| **Shot #** | **N220111-2** | **N240102-1** | **N220214-2** | **N220607-3** | **N240103-1** | **N221121-1** | **N220703-1** | **N240104-1** | **N230105-2** | **N230207-1** |
|---|---|---|---|---|---|---|---|---|---|---|
| **Hohlraum Type** | FIG. 4 a) | FIG. 4 a) | FIG. 4 b) | FIG. 4 b) | FIG. 4 c) | FIG. 4 c) | FIG. 4 d) | FIG. 4 d) | FIG. 4 d) | FIG. 4 d) |
| **LEH diameter [mm]** | 4.60 | 4.60 | 3.11 | 3.11 | 3.11 | 3.11 | 3.11 | 3.11 | 3.11 | 3.11 |
| **LEH Hardware** | None | None | None | None | LEH Mem. Only | LEH Mem. & Storm Wind. | LEH Mem. & Storm Wind. | LEH Mem. & Storm Wind. | LEH Mem. & Storm Wind. | LEH Mem. & Storm Wind. |
| **Hohlraum Fill [mg cm$^{-3}$]** | Vacuum | Vacuum | Vacuum | Vacuum | Vacuum | Vacuum | Vacuum | He 0.02 mg cm$^{-3}$ | He 0.03 mg cm$^{-3}$ | He 0.03 mg cm$^{-3}$ |
| **Hohlraum Dopant Patches** | Top Mn Bottom Fe | Top Mn Bottom Fe | Top Mn Bottom Fe | None | None | None | None | None | None | None |
| **Capsule** | No Capsule | No Capsule | No Capsule | No Capsule | No Capsule | No Capsule | HDC capsule | HDC capsule | HDC capsule | HDC capsule |
| **Cap. Fill** | NA | NA | NA | NA | NA | NA | None | $D^3He$ 4.0 mg cm$^{-3}$ | $D^3He$ 4.0 mg cm$^{-3}$ | $D^3He$ 4.0 mg cm$^{-3}$ |
| **Pulse shape** | FIG. 5 a) 2ns square | FIG. 5 a) 2ns square | FIG. 5 a) 2ns square | FIG. 5 a) 2ns square | FIG. 5 b) 2ns square w/ toe & foot | FIG. 5 b) 2ns square w/ toe & foot | FIG. 5 b) 2ns square w/ toe & foot | FIG. 5 b) 2ns square w/ toe & foot | FIG. 5 b) 2ns square w/ toe & foot | FIG. 5 c) Realistic pulse |
| **Target Temperature** | Warm | Warm | Warm | Warm | Warm | Warm | Warm | Cryo | Cryo | Cryo |
| **Laser Drive Energy [kJ]** | 912±18 | 903±18 | 893±18 | 900±18 | 936±19 | 909±18 | 922±18 | 925±20 | 885±18 | 942±19 |
| **Dante x-ray data parameters specified at peak – Experimental & Simulation** | | | | | | | | | | |
| **Dante-1** | | | | | | | | | | |
| $I_1^E$ [TW sr$^{-1}$] | 27.0±1.3 | 26.6±1.3 | 15.4±0.8 | 16.7±0.8 | 16.4±0.8 | 16.7±0.8 | 14.6±0.7 | 13.3±0.7 | 13.9±0.7 | 14.5±0.7 |
| $I_1^S$ [TW sr$^{-1}$] (LHT) | 29.5 | 29.5 | 16.5 | 17.0 | 17.4 | 17.4 | 15.0 | 15.4 | 14.6 | 14.8 |
| $I_1^E/I_1^S - 1$ [%] | -0.09±0.05 | -0.10±0.05 | -0.06±0.05 | -0.01±0.05 | -0.05±0.05 | -0.04±0.05 | -0.03±0.05 | -0.14±0.05 | -0.05±0.05 | -0.02±0.05 |
| $T_{R1}^E$ [eV] | 281±4 | 280±3 | 297±4 | 303±4 | 302±4 | 303±4 | 293±4 | 286±4 | 289±4 | 292±4 |
| $T_{R1}^S$ [eV] (LHT) | 290 | 287 | 302 | 302 | 306 | 304 | 294 | 297 | 294 | 294 |
| $I_{M1}^E$[TWsr$^{-1}$] | No Data | 3.0±0.3 | 1.9±0.2 | 2.1±0.2 | 2.0±0.2 | 2.0±0.2 | 1.8±0.2 | 1.4±0.1 | 1.7±0.2 | 1.7±0.2 |
| $I_{M1}^S$[TWsr$^{-1}$] (LHT) | No Data | 4.7 | 2.7 | 2.8 | 3.0 | 2.9 | 2.5 | 2.6 | 2.4 | 2.4 |
| $I_{M1}^E/I_{M1}^S - 1$ | No Data | -0.37±0.06 | -0.29±0.06 | -0.26±0.07 | -0.33±0.06 | -0.34±0.06 | -0.29±0.06 | -0.46±0.06 | -0.30±0.06 | -0.28±0.06 |
| **Dante-2** | | | | | | | | | | |
| $I_2^E$ [TW sr$^{-1}$] | 9.5±0.9 | 9.5±1.0 | 5.7±0.6 | 5.8±0.6 | 6.7±0.7 | 4.9±0.5 | 4.2±0.4 | 4.9±0.5 | 4.7±0.5 | 5.0±0.5 |
| $I_2^E$ [TW sr$^{-1}$] Corrected* | NA | NA | NA | NA | NA | 6.7±0.7* | 5.8±0.6* | 6.8±0.7* | 6.5±0.6* | 6.8±0.7* |
| | | | | | | | | | | |
| $I_2^S$ [TW sr$^{-1}$] (LHT) | 11.9 | 11.9 | 7.0 | 7.0 | 6.1 | 6.2 | 5.2 | 5.2 | 6.2 | 6.3 |
| $I_2^E/I_2^S - 1$ | -0.21±0.08 | -0.20±0.08 | -0.18±0.08 | -0.16±0.08 | +0.10±0.11 | +0.09±0.11* | +0.12±0.11* | +0.30±0.13* | +0.04±0.10* | +0.08±0.11* |
| $T_{R2}^E$ [eV] | 251±6 | 251±6 | 269±7 | 271±7 | 281±7 | 281±7* | 270±7* | 281±7* | 278±7* | 281±7* |
| $T_{R2}^S$ [eV] (LHT) | 266 | 266 | 283 | 283 | 274 | 275 | 263 | 263 | 275 | 276 |
| **Gold bubble plasma conditions data from analysis of NSS Au l-shell Spectroscopy Data– Experimental & Simulation** | | | | | | | | | | |
| Exp Wall $T_e$ (NSS) [keV] | NA | NA | NA | 3.95±0.2 | NA | NA | 3.6±0.2 | NA | 3.8±0.2 | 3.8±0.2 |
| Sim. Wall $T_e$ [keV] (LHT) | NA | NA | NA | NA | NA | NA | 2.92 | NA | 2.87 | 3.02 |
| Exp Wall $\overline{Z}$ (NSS) | NA | NA | NA | 54.82±0.12 | NA | NA | 54.15±0.21 | NA | 54.18±0.13 | 54.79±0.12 |
| Sim. Wall $\overline{Z}$(LHT) | NA | NA | NA | 52.63 | NA | NA | 52.68 | NA | 52.63 | 52.56 |
| **Bang time data – Experimental & Simulation** | | | | | | | | | | |
| Bang-time (Spider)[ns] | NA | NA | NA | NA | NA | NA | NA | 5.38±0.04 | >5.2ns† | NA |
| Bang-time (2DConA)[ns] | NA | NA | NA | NA | NA | NA | NA | 5.47±0.08 | NA | NA |
| Bang-Time (PTOF)[ns] | NA | NA | NA | NA | NA | NA | NA | 5.34±0.11 | NA | 7.35+0.5-0.15 |
| Bang-time (LHT) [ns] | NA | NA | NA | NA | NA | NA | NA | 4.79 | 4.4 | 7.03 |
| Bang-time discrep. [ns] | NA | NA | NA | NA | NA | NA | NA | 0.63±0.05 | >0.8 | 0.33+0.5-0.15 |

Table 2 Table of key experimental results and predictions of simulations using the Lasnex Hohlraum Template. * These Dante-2 flux measurements include a 27% correction for occlusion of the line of sight by the LEH hardware. †Only rising edge of "bang" emission was observed for this measurement.

The drive spectrum can be divided into a thermal portion which is modeled as a blackbody spectrum and an NLTE gaussian “bump” corresponding to M-band transitions in the higher temperature and lower density ablated gold wall plasma[73]. A dedicated 2 - 4 keV filter channel has recently been added to the Dante-1 line of sight[45]. This channel has an approximately flat response to x-ray power over its spectral band. Within the parameter space of our experiments the uncertainty in the spectral intensity measured by this channel is ~ 9% [45]. The flux measured by this channel at peak radiant intensity is tabulated as $I_M$ in Table 2, reflecting the fact that this channel is sensitive to the gold M-band emission. It should be noted that this channel is also sensitive to the high energy tail of the gold thermal emission.

The experimental drive metrics are compared with similar metrics determined via analysis of post-shot Lasnex LHT simulations. These simulations use the “as-shot” measured laser drive power, therefore providing a direct comparison to the measurements. The simulated plasma maps are post-processed to calculate the $dI/dE$ in the direction of the two Dante diagnostics. These data are then integrated over energy to reproduce the experimental metrics described above. Simulations and experiments are compared using the metric $I^E/I^S - 1$. Positive values indicate experimental measures are higher than expected based on simulation, while negative values reflect experimental measurements that are lower than expected.

For experiments using a storm window, the Dante-2 view of the LEH is partially obscured by the storm window mounting hardware (see FIG. 7 b)). This mounting hardware is not fully modeled in simulations (see FIG. 3). The effect of this obscuration was investigated both through calculation and experiment, resulting in the adoption of a 27 % correction factor which was applied to the relevant $I_2^E$ measurements in Table 2. Section V.B.5 provides more details on why these corrections to our experimental measurements are required, and how the size of the correction was assessed.

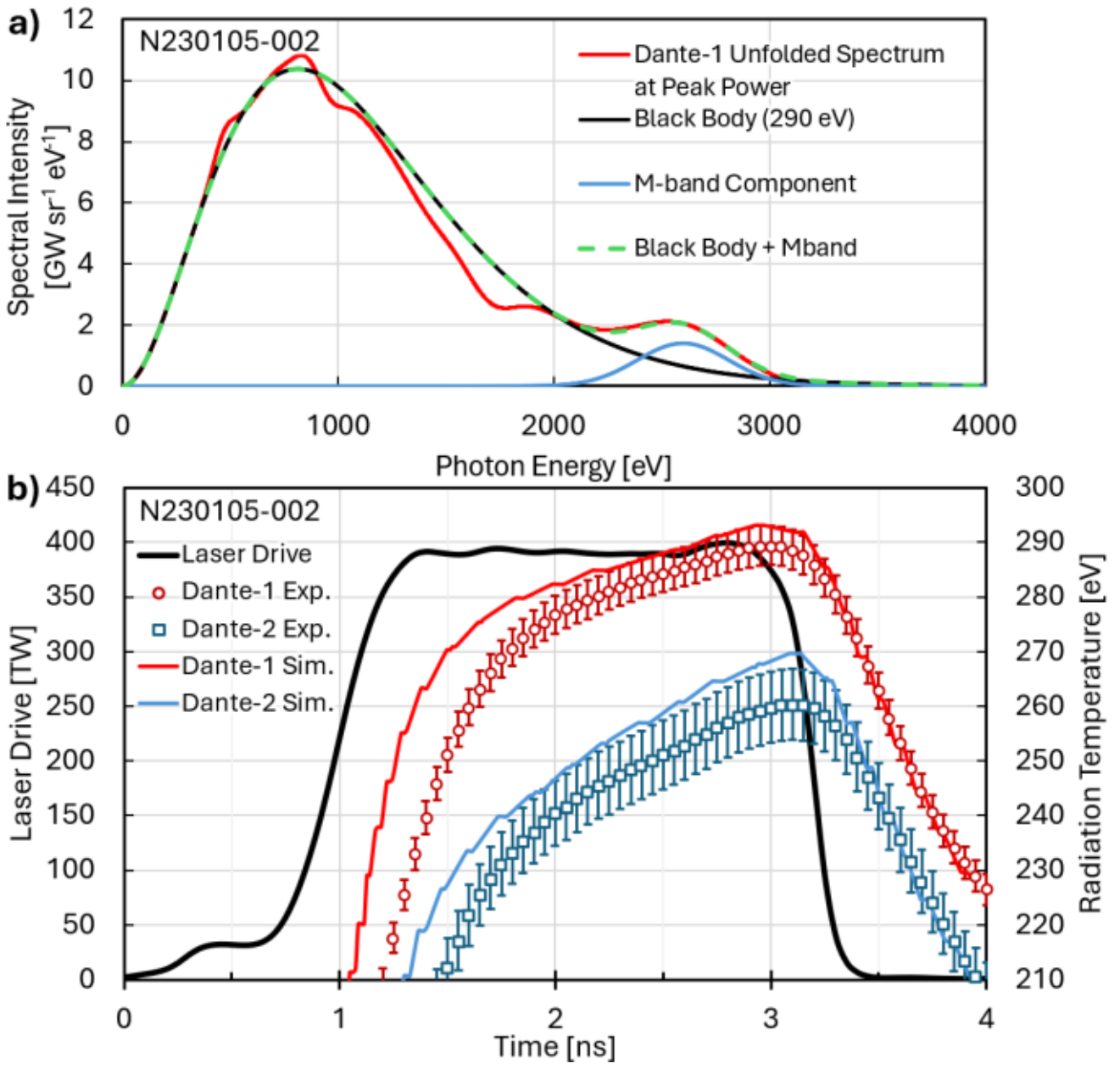

FIG. 6 Example Dante data from N230105-002. a) Dante-1 spectrum measured at peak power. The spectrum is modeled as a blackbody with an additional gaussian bump to model the M-band emission. b) Similar analysis is performed at each time step to produce a time-resolved measurement of the effective radiation temperature of the LEH from the Dante line of sight.

Experiment N240104-1 is clearly an outlier so far as the Dante radiation temperature measurements are concerned. Using the $I^E/I^S - 1$ metrics, we see the Dante 1 measurement is 14% lower than simulation compared to an average of ~6% for the other experiments, while Dante-2 is 30% high compared to an average of ~9% across the other shots. The m-band emission is 46% low compared to an average of ~30% across the other shots. This discrepancy is due to this experiment using a “2DConA” capsule backlighter diagnostic[74] to facilitate a more accurate measurement of the capsule implosion trajectory. A pair of laser quads from the upper and lower outer 50° cones were redirected from the hohlraum to drive a backlighter foil. The power of the remaining beams in the 50° cone were increased to compensate for these missing quads but no changes were made to the pointing of the remaining beams to reestablish drive symmetry. Azimuthal asymmetries in the drive power cannot be captured in our 2D Lasnex simulations, nor can we include emission from the backlighter. The Dante-2 spectrometer has a direct line of sight to the backlighter, explaining why its radiant intensity measurement is so high compared to the other experiments in the campaign. While Dante1 also sees the backlighter x-ray source, the beams used to drive the backlighter happen to be the ones that normally heat the portion of the hohlraum wall in its line of sight and so overall it sees lower peak radiant intensity and much lower M-band emission. For this reason, the Dante data from this experiment has been disregarded in our analysis and discussion and excluded from all plots and averages discussed in this section.

### *2. Comparison of peak x-ray power between experiment and simulation*

Trends in x-ray drive can be seen in the measured $T_{R1}^E$ and simulated $T_{R1}^S$ tabulated in Table 2 and plotted in FIG. 8 a). These

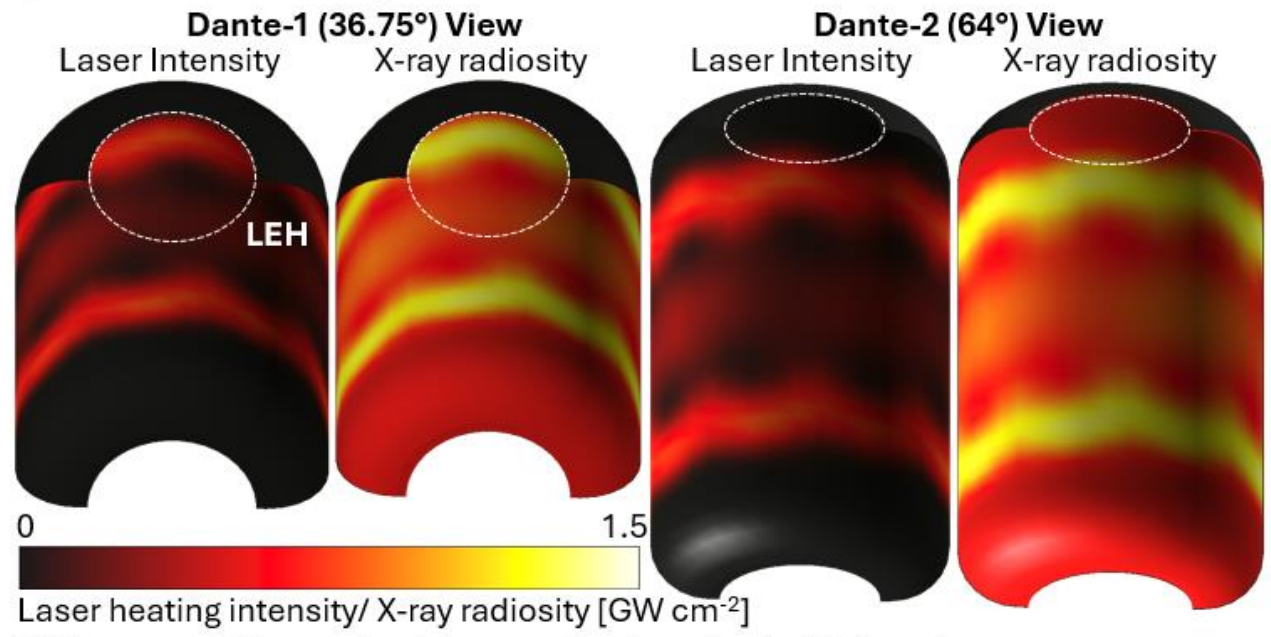

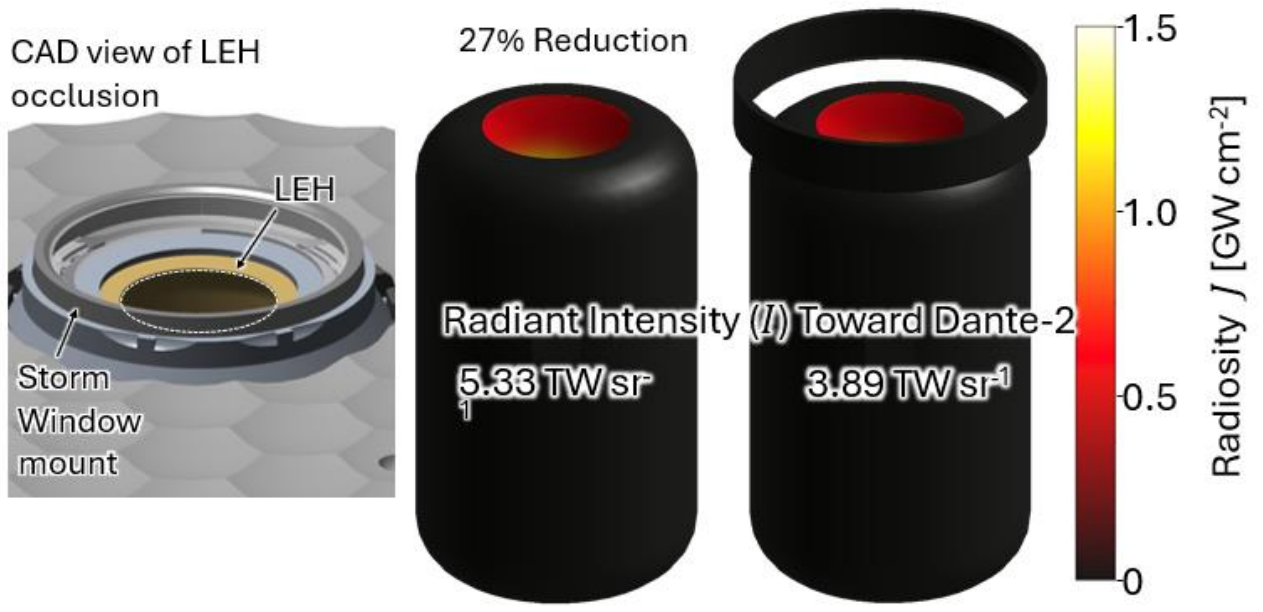

FIG. 7 a) Dante views of the interior of the hohlraum. The near half of the hohlraums are cut away to provide a clear view of the laser heating and x-ray radiosity patterns. White dashed circles show the portion of the wall viewed by each diagnostic. Dante-1 views the outer ring of laser spots and inner wall area. Dante-2 mainly views the unheated portion of the wall above the outer ring beam spots. Calculated using VisRad assuming constant laser power equal to the 400 TW max used in our experiment. X-ray conversion efficiency was set to 0.95. The hohlraum wall albedo parameter was adjusted to 0.77 to match experimental measurements of peak radiant intensity along the Dante-1 & Dante-2 lines of sight. b) Occlusion of the LEH along the Dante-2 line of sight by the storm window is calculated to reduce measured radiant intensity by 27%. The storm window predominately blocks the region of brightest emission, at the edge of the inner beam ring.

trends can be understood in the context of simple hohlraum performance model summarized in section II. The hohlraums with a 4.6 mm Ø LEH reached a peak $T_{R1}^{E} \sim 280$ eV. When the LEH size decreased to 3.1 mm $T_{R1}^{E}$ increased to $\sim 300$ eV, consistent with reduced x-ray losses through the smaller LEH. No significant change in peak $T_{R1}^{E}$ was observed when the LEH hardware was added. The addition of a capsule led to a further reduction in peak $T_{R1}^{E}$, to $\sim 290$ eV, consistent with the additional hohlraum losses expected due to absorption of x-rays by the capsule, and in excellent agreement with the simple power balance model prediction presented in section II of 295eV. No significant additional reduction in peak $T_{R1}^{E}$ was seen when the hohlraum gas fill was added. These trends are qualitatively as expected based on our analytical estimates, with no obvious

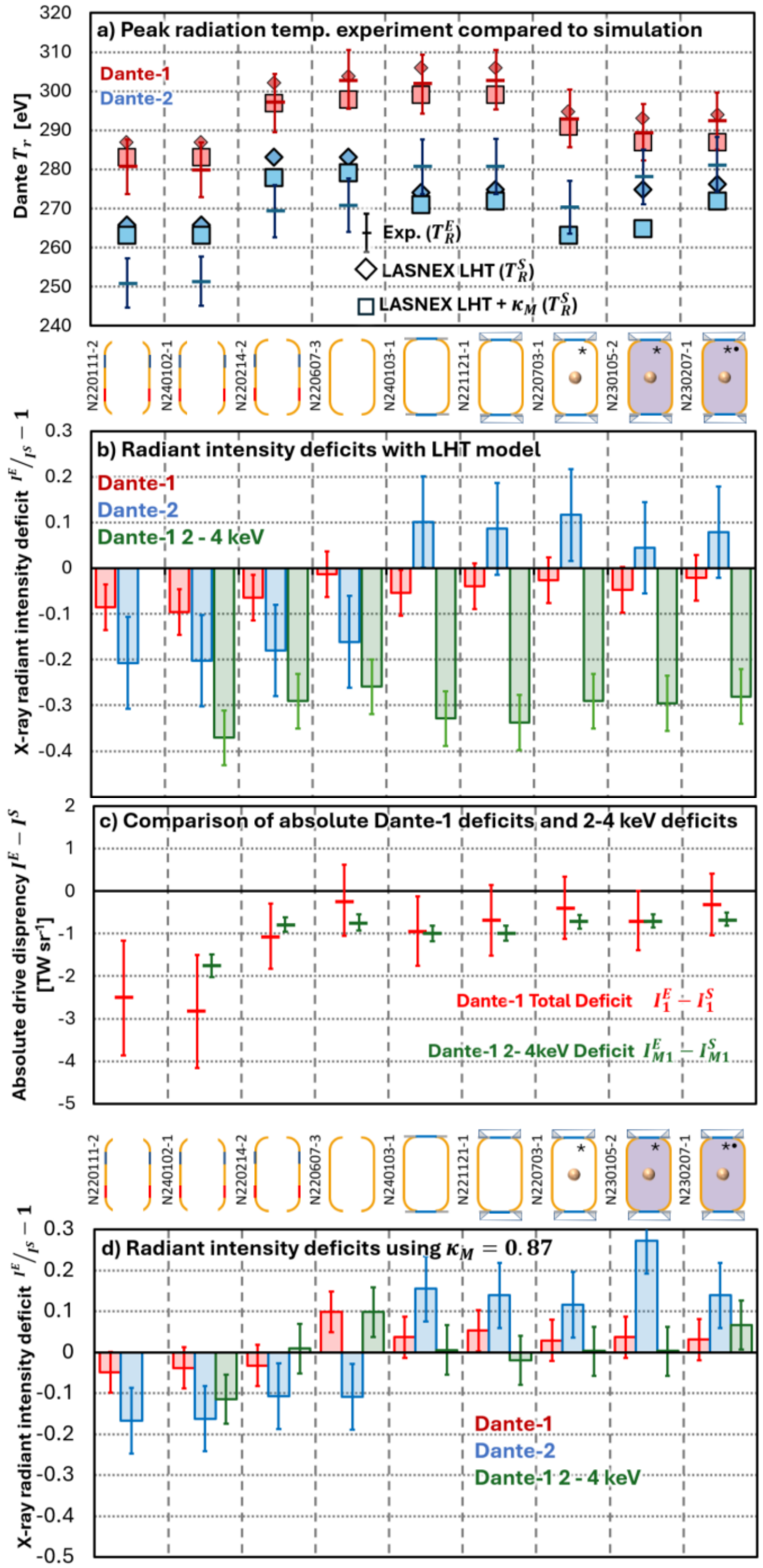

FIG. 8 a) Plots of the peak drive temperature $T_r$ measured by Dante and simulated using Lasnex b) Fractional deficits in experimentally measured radiant intensities $I$. c) Comparison of total radiant intensity deficit for Dante-1 and the 2 – 4 keV channel d) Radiant intensity deficits when using use a $\kappa_M = 8.7$ opacity multiplier to suppress high energy photon emission. Notes: * Dante-2 $I_2^E$ corrected for occlusion (see V.B.5). • Laser drive used the realistic laser pulse (FIG. 5 c)).

unexpected drive losses as target features are added. Identical trends are seen in the Lasnex LHT simulated radiation temperatures for Dante-1 ($T_{R1}^{S}$), albeit at slightly higher temperatures across the board.

The Dante-2 measurements ($T_{R2}^{E}$) follow a similar trend to Dante-1, but at lower radiation temperatures, as the portion of the wall viewed by Dante-2 is not directly laser heated (see FIG. 7 a)). Unlike for Dante-1, experiments show a clear additional step up in $T_{R2}^{E}$ of about 10 eV when moving from the “simple” targets to the more complex targets which include the LEH hardware. This effect is seen most clearly by comparing experiment N220607-3 with N240103-1, as former experiment did not include the storm window assembly and therefore ought to have an un-occluded Dante-2 line of sight into the hohlraum. The $T_{R2}^{S}$ values simulated using the LHT interestingly show the opposite trend, with radiant intensity falling when the LEH hardware is added. For the hohlraums without an LEH membrane $T_{R2}^{S}$ is systematically ~15 eV higher than $T_{R2}^{E}$ while for the hohlraums with an LEH membrane $T_{R2}^{S}$is systematically ~ 5 eV lower than $T_{R2}^{E}$. The underlying cause is not yet understood but may be related to inaccuracies in the modeling of the LEH hardware. Another consistent difference between these sets of experiments is that targets with an LEH membrane also used a small foot on the drive pulse (see FIG. 5).

The discrepancy in the experimentally measured radiant intensity is parameterized as $\mathrm{I^E/I^S} - 1$. This is also tabulated in Table 2 for each Dante line of sight. The uncertainty in the experimental measurement of $I^E$ has been propagated to this ratio. A plot of these deficits is provided in FIG. 8 b). Overall, the individual $I_1^E$ measurements are in reasonable agreement with the simulated $I_1^S$, with almost all the $I_1^S$ falling within the uncertainty band of the corresponding measurement, however the average of $\mathrm{I_1^E/I_1^S} - 1$ across the campaign is - 0.06±0.01, indicating measured drive is systematically ~6% lower than predicted by LHT simulations. Note that this argument relies on the uncertainties in the individual Dante measurements being random. This assumption appears reasonable given that the error in each individual measurement results from a combination of many different independent error sources (filter transmissions, aperture sizes, diode sensitivities, cable losses, attenuator losses for each of 18 channels). Furthermore, many of these components were exchanged or modified over the course of the campaign, which would tend to disturb a consistent systematic error. We cannot, however, fully exclude that a shift in the mean 5% level does not result from systematic error. This ~ 5% discrepancy in total x-ray emission is not inconsistent with the level of agreement between simulation and experiments claimed in previous studies[42,43,75].

For Dante-2, the agreement between simulation and experiments for the “complex” two-part hohlraums is reasonably good. The mean experimental deficit $\mathrm{I_2^E/I_2^S} - 1$ across these complex target shots is 0.09±0.1, i.e. experimental measurements are consistently 9% higher than predicted. Conversely, for the simpler, one-part hohlraums, measurements are consistently ~ 20 % lower than predicted using the LHT.

Adding a capsule to the target did not have a significant impact on the agreement between experiment and simulation for either Dante channel. This is seen in the similarities in the x-ray discrepancies for experiments N221121-1 & N220703-1. The only change between these experiments was the addition of a capsule. Similarly, we see little change with the addition of a gas fill when comparing experiments N220703-1 & N230105-2.

### 3. Comparison of the Dante-1 2 – 4 keV emission

The spectral intensity deficit for the Dante-1 dedicated 2 - 4 keV filter channel $I_{1M}^{E}/I_{1M}^{S} - 1$ is tabulated in Table 2, and plotted in FIG. 8 b). The radiant intensity measured by this channel is on average 31% lower than simulation across the suite of experiments, with a standard deviation of 3%. This discrepancy in higher energy photon emission is much larger than the ~9% measurement uncertainty. These measurements are consistent with an overprediction of the M-band emission from the gold bubble in simulations.

Inspecting FIG. 8 b), there are some trends in $I_{M1}^{E}/I_{M1}^{S}-1$ with design complexity. The discrepancy for the hohlraum with a 4.6 mm LEH is larger than that for the 3.1 mm LEH simple hohlraums. The discrepancy increases a little with the addition of the LEH hardware, and then decreases a little again with the addition of the capsule, indicating that this measurement is sensitive to the detailed hohlraum dynamics.

There is a very good agreement between the absolute radiant intensity deficit in the 2 – 4 keV band and the total radiant intensity deficit measured across the campaign. This is illustrated on a shot-by-shot basis in FIG. 8 c). Overall, across the targets using a 3.1 mm LEH, the average absolute discrepancy in peak M-band radiant intensity is $\left|I_{1M}^{E}-I_{1M}^{S}\right|$ = -0.9±0.1 TW, closely matching the average absolute total Dante-1 drive discrepancy $\left|I_{1}^{E}-I_{1}^{S}\right|$ = -0.9± 0.3 TW.

### *4. Discussion*

Using Dante-1, our experiments measure ~ 30 % lower emission in the 2 – 4 keV photon energy band than is predicted by simulation. Over the complete ensemble of experiments total emission from hohlraums measured by Dante-1 is consistently ~ 5 % lower than predicted. As is illustrated in FIG. 8 c), this 5 % lower drive can be completely explained by the measured deficit in the 2 - 4 keV range emission. The deficit in this high energy photon range is consistent with an overprediction of the M-band emission from the gold bubble in the LHT simulations. This observation supports the hypothesis that errors in the NLTE physics modeling in our radiation hydrodynamics codes are responsible for a significant portion of the historical bang-time discrepancy. Further discussion of the discrepancy in M-band emission and modifications to our simulations to better match experiment are presented in section VI, along with an assessment of the overall effect this has on the bang time discrepancy.

For Dante-2 the experimentally measured total drive is on average ~9 % higher than simulation in experiments that include LEH hardware, but ~15-20 % lower compared to simulation for experiments without the LEH hardware. There are several possible explanations for the change in agreement when the LEH hardware is added.

- The 27% LEH hardware occlusion factor correction is too large.
- LEH closure in targets without LEH membranes is underestimated.
- The line of sight to the Dante-2 diagnostic is somehow blocked in the vacuum hohlraum experiments, perhaps by material ablated from the top of the hohlraum
- Presence of the LEH membrane modifies beam propagation into the hohlraum, reducing portion of the laser spots viewed by Dante-2
- Some difference is introduced by the low power foot used to precondition the LEH membrane.

We have not been able to determine the underlying cause of this discrepancy based on the available data. Dante-1 has a better line of sight into the target with less potential for complications and no requirement for corrections, so we have weighed its measurements more heavily in the development of our conclusions.

The absence of changes in the agreement between experiment and simulation when the capsule and hohlraum gas fill are added indicate that mix in the hohlraum is not a dominant energy loss mechanism.

### *5. Effect of line-of-sight obscuration for Dante-2 measurements*

While a model of the LEH hardware was included in the relevant Lasnex simulations, that model does not include the outer aluminum storm window retaining ring (see comparisons or real and simulated hardware provided in FIG. 3). The aluminum retaining ring dominates the attenuation of x-rays towards Dante-2; the Rexolite storm window mount that is included in the simulations is mostly transparent to this radiation. Without the retaining ring the effects of the LEH occlusion cannot be accurately calculated from the simulated plasma maps. Acknowledging this limitation, the LEH hardware was cropped from the simulation plasma maps during calculation of $I^{S}$. The values of $I_{2}^{S}$ tabulated in Table 2 therefore do not include any hardware occlusion effect. To allow direct comparison with experiment, the measurements of $I_{2}^{E}$ in Table 2 were corrected using the 27% obscuration factor described in this section.

An initial simple geometric calculation showed that ~20% of the area of Dante-2 view of the hohlraum wall through the LEH would be obscured by the storm window mounting hardware. This estimate does not account for variations in the radiosity ($J\ [TWcm^{-2}]$) of the hohlraum wall. A more involved calculation using the 3D view factor code VisRad[76] produced a larger estimate of 27% for the overall fraction of radiant intensity blocked by the storm window mounting washer. The outputs generated in this calculation are presented in FIG. 7 b). The hohlraum target was configured to match the initial size and shape of that used in our experiments and the laser model was configured to match our experimental design. The laser drive was set to match the peak power used in our experiments. The hohlraum was modeled in VisRad using the setup described in FIG. 7 a). Occlusion of the Dante-2 line of sight was mocked up by placing an annular ring around and above the hohlraum LEH to match the occlusion expected to be provided by the storm window retaining ring. The radiant intensity from each hohlraum was found by calculating the irradiance $E$ [W m$^{-2}$] on a small surface element specified at a 6 m distance along that Dante-2 measurement line of sight, facing the hohlraum. This irradiance was converted to a radiant intensity based on the standoff distance. The occlusion factor was calculated as the ratio of these two radiant intensities.

Inspecting the maps of the hohlraum radiosity $J$ [W cm-2] in FIG. 7 b), for the unobscured hohlraum $J$ is clearly highest at the lower lip of the hohlraum. This enhanced emission comes from the edge of the ring of outer beam spots on the hohlraum wall, shown more clearly in FIG. 7 a). The storm window washer predominantly blocks the view of this higher emission portion of the hohlraum wall, thus explaining why a larger reduction in radiant intensity is seen than the geometric estimate first suggested.

An experiment was conducted to verify these estimates. Experiment N240103-1 was a close repeat of experiment N221121-1. The targets were identical apart from the removal of the storm window hardware for N240103-1. The peak $I_{2}^{E}$ measured without the storm window hardware was 6.7±0.7 TW sr$^{-1}$, compared to 4.9±0.5 TW sr$^{-1}$ for the experiment with the storm window hardware. This reduction of 27% in measured radiant intensity is in excellent agreement with our VisRad estimates.

## C. LEH Size measurements

The GLEH[27,28] imager was used to observe the evolution of the hohlraum LEHs using the lower static x-ray imager (SXI)[25,47] line of sight (see FIG. 2). SXI Lower views the hohlraum LEH from an angle 19° from the hohlraum axis and provides measurements of both the LEH size as a function of time and the location of laser spots on the hohlraum wall. Images are recorded using a multi-frame time-gated hybrid CMOS image sensor.

The method used to determine the LEH diameter from the GLEH images is described in detail section II.B.2 of reference [27]. Typical uncertainties for these measurements are 50 µm. The time resolved LEH size measurement was made on a total of four experiments, distributed across the campaign. The experimental measurements are compared to simulated LEH sizes determined through post-processing of plasma maps. The method used to extract

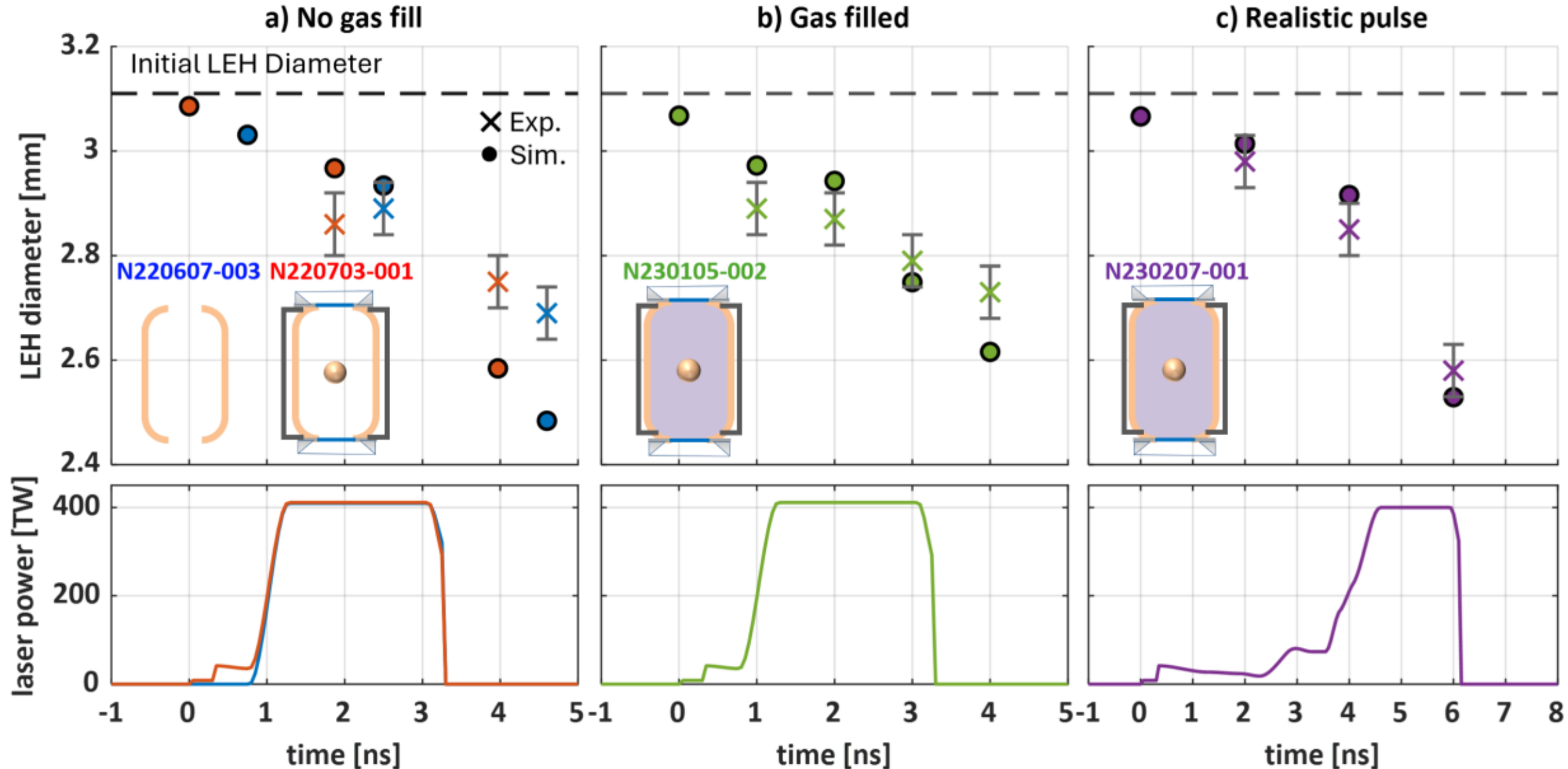

FIG. 10 LEH diameter experimental data compared to LASNEX LHT simulations across the campaign using GLEH. Simulations underestimate LEH closure at early times and overestimate close at late times. Reasonable agreement is seen at the time of peak x-ray emission, at the end of the laser pulse.

LEH diameter from the simulations is conceptually similar to that used for the experimental data[27], allowing us to make direct comparisons between measurement and simulation.

FIG. 10 presents a summary of the results of these measurements. The first panel (a) compares two vacuum hohlraum experiments. Data from the first shot has been time-shifted by 0.75 ns to align the high-power parts of the two pulses temporally. The second panel (b) shows a similar measurement for a gas filled hohlraum and the final panel (c) shows the data for a gas filled hohlraum using the more realistic laser pulse.

At early times the experimentally measured size of the LEH is found to be smaller than predicted by simulations. Discrepancies of up to 100 μm are seen during the highest power part of the drive pulse. As discussed in section II.3, this discrepancy would lead to an ~2% reduction in simulated coupling efficiency to the capsule and an ~4% increase in simulated radiative emission from the hohlraum. Towards the end of the drive and after the drive ends the discrepancy in LEH size decreases and then reverses, with the LEH size measured after the end of the drive being larger than in our simulations.

The main parameter that we compared for our x-ray measurements is the peak radiant intensity of x-ray emission from the hohlraum. The hohlraum temperature rises throughout the flat part of the laser pulse due to the increasing hohlraum wall albedo associated with deeper penetration of the Marshak wave[30]. This means that our comparisons of peak radiant intensity are made at the time just before the laser turns off. Returning to the data plotted in FIG. 10, we see that generally the experimentally measured and simulated LEH diameters agree much more closely at this time, and therefore LEH size discrepancies earlier in time should not have a significant impact on this comparison.

The very limited amount of data currently available does not allow us to draw strong conclusions regarding LEH closure. Agreement between experiment and simulation at peak radiant intensity appears reasonably good but there is evidence that simulations both underestimate closure at early times and overestimate closure speed once the laser drive turns off. This may help to explain why x-ray drive rises earlier in simulations and then falls more quickly after the lasers turn off. The LEH size discrepancy does not provide a means to explain the bang-time discrepancy. A smaller LEH early in time in our experiments ought only to increase coupling of x-ray energy to the capsule due to reduced losses through the LEH. We therefore conclude that errors in the modeling of LEH closure are unlikely to be able to explain the bang-time deficit.

## D. Gold bubble temperature and ionization state measurements

The gold bubble temperature and ionization state were diagnosed using the NIF Survey Spectrometer (NSS)[46]. This is an x-ray spectrometer that observes the target from an angle of 36.75° from the hohlraum axis, i.e. along a similar line of sight to the Dante-1 line of sight illustrated in FIG. 7 a). This view provides a clear line of sight to the ring of outer beam spots that drive the formation of the gold bubble region. NSS records time-integrated spectra over the 7.5 - 13.5 keV energy range, providing measurements of inner-shell

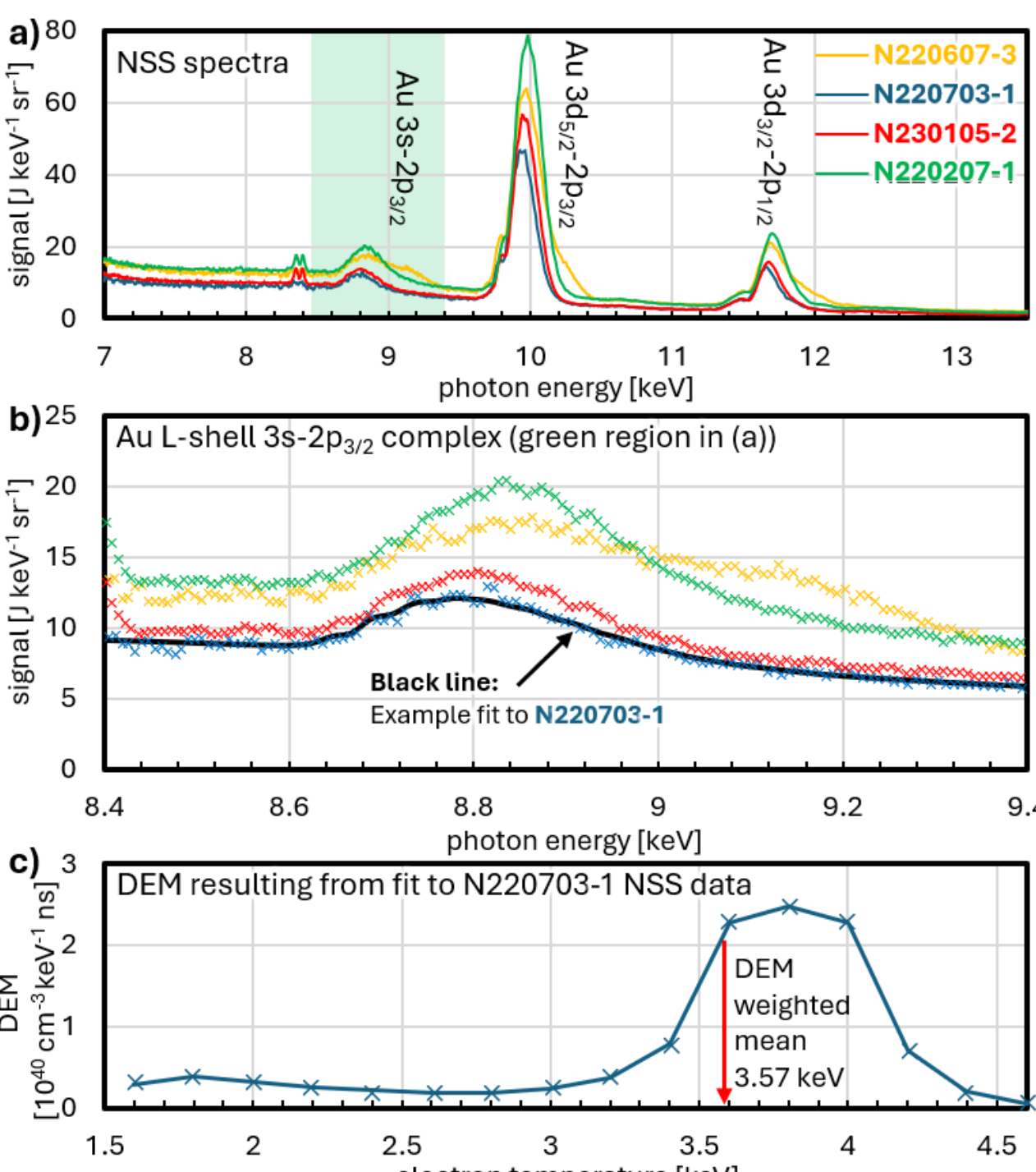

FIG. 9 Examples of NSS data from the Build-A-Hohlraum campaign. a) spectra measured in a range of experiments. b) zoom in to the 3s-$2p_{3/2}$ Au L-shell complex (green shaded region in a)). An example fit to the data for shot N220703 is provided. c) DEM corresponding to the fit provided in b), illustrating the DEM weighted average temperature determined from this fit.

Au L-shell emission from M-shell ions. The raw spectra recorded in each experiment by NSS are plotted in FIG. 9 a).

The spectral data are used to infer the Au bubble temperature and ionization state. The 8.4 – 9.4 keV Au L-shell 3s-2p complex, indicated by the green shaded region in FIG. 9 a) and expanded on in FIG. 9 b), is fitted using a genetic algorithm combined with the atomic kinetics code SCRAM[77]. An example fit is shown for experiment N220703-1. Fitting the spectral data results in a Differential Emission Measurement (DEM) of plasma in the diagnostic line of sight. The DEM is a density-squared and volume weighted temperature distribution of the plasma contributing to the signal[78].

$$DEM(T) = n_e^2 \frac{dV}{dT} \qquad (13)$$

This definition is appropriate for two-body emission processes, such as Au L-shell emission, which is driven by inner-shell collisional excitation of M-shell ions. The DEM corresponding to the fit shown in FIG. 9 b) is plotted in FIG. 9 c). The gold bubble temperatures is parameterized as the DEM-weighted average. An ionization state corresponding to each temperature bin in the DEM is determined using SCRAM, facilitating the calculation of a DEM weighted mean $\bar{Z}$. A much more detailed discussion of the analysis methods summarized here will form the focus of a separate, upcoming paper[79].

The DEM-weighted average $T_e$ and $\bar{Z}$ for three different targets are plotted in FIG. 11. These experimental data are compared to values determined from simulations. For the simulated measurements the DEM is calculated directly from the Lasnex plasma map and the value plotted is the weighted average of that DEM. It is clear from FIG. 11 that the experimentally infered gold bubble temperatures are significantly higher than those predicted by the Lasnex Hohlraum Template (LHT) simulations. The average experimentally infered temperature is ~3.7 keV compared to an average simulated temperature of ~2.7 keV, a ~1 keV temperature difference. Simulations run with the Lasnex MHD package enabled showed slightly closer agreement, with experiment, raising the plasma temperature significantly, but were insufficient to account for the full discrepancy. Corresponding discrepancies are observed in the Au average ionization state. The experimentally infered $\bar{Z}$ ~54.5, compared to ~52.5 in the LHT simulations.

The gold bubble temperature discrepancy provides further evidence of a significant issue with the NLTE modeling of the hohlraum wall emission in the Lasnex LHT. These observations are qualitatively consistent with the discrepancy in the 2 – 4 keV photon emission discussed in section V.B – a reduced effective opacity of this plasma would lead to reduced heat loss and a higher temperature in the bubble. Similar temperature discrepancies were also observed in recently published ViewFactor experiments[32]. This result and approaches to approve agreement between experiments and modeling are further discussed in section VI.

Hohlraum wall mix with the gas fill or ablated capsule plasma have also been predicted to drive increases in the gold bubble temperature[29]. This hypothesis can be excluded as the driving force behind the temperature discrepancy we see in the BAH experiments as we see these elevated temperatures across all targets, including the simple vacuum hohlraums where there should be no mix effects (see FIG. 11).

## E. Bang-time measurements

Fueled capsules were fielded in three experiments, as summarized in Table 2. For these experiments the bang-time was measured to provide a direct comparison to the x-ray drive deficit as measured by Dante. The methods used to determine the bang-time varied between shots based on diagnostic availability.

For experiment N230105-2 the bang-time was determined using the SPIDER[80] x-ray streak camera. Spider is typically the diagnostic of record for measurements of x-ray bang-time. The target used in this experiment was a full ID-ICF platform target, using a 2-part hohlraum with the full TMP, LEH hardware set, capsule and cryogenic hohlraum gas fill. The laser pulse used to drive the target was the ~2 ns square pulse with foot and toe, plotted in FIG. 5 b). A diagnostic timing error in this experiment meant that the bang-time was not directly observed. We see the initial rise in signal associated with the start of capsule emission but do not see the peak of the emission. This limits this measurement to placing a lower limit on the bang-time > 5.2 ns. Our LASNEX LHT simulation predicted a bang-time of 4.4 ns, suggesting a bang-time discrepancy of >0.8ns. This discrepancy is larger than the ~400 ps discrepancy seen in historical ID-ICF shots (see FIG. 1), but we note that the 2 ns square drive used in these experiments is quite different from a typical ICF drive.

Experiment N240104-1 fielded a very similar target, but with a modified laser drive setup to accommodate a backlighter foil to facilitate a 2D x-ray radiography measurement (2DConA)[74] of the imploding capsule. Although two outer quads were redirected to drive the backlighter, the remaining outer quads were increased in power and energy to compensate. No repointing of the remaining beams was carried out to mitigate the low-mode asymmetry introduced by the dropped beams. In this experiment bang-time was measured at 5.38±0.04 ns by both the 2DConA and Proton Time of Flight (PTOF)[81] measurements, ~0.6 ns later then in the LHT simulation.

Finally, experiment N230207-1 again used a similar target but this time used the more realistic multi-tiered laser pulse shape (FIG. 5 c)). Spider was unavailable for this experiment and bang-time was determined from a PTOF measurement. This measured 7.35+0.5-0.15 ns. The asymmetric error bars reflect issues with the later part of the signal due to measurement noise. A bang-time of 7.35 ns would be 0.33 ns late with respect to simulation, in good agreement with the historical ID-ICF data which indicated a mean error of 400 ps (FIG. 1). Using an opacity multiplier in simulations to reduce emission of higher energy photons improves agreement between simulation and experiment. This is discussed in more detail in section VI.

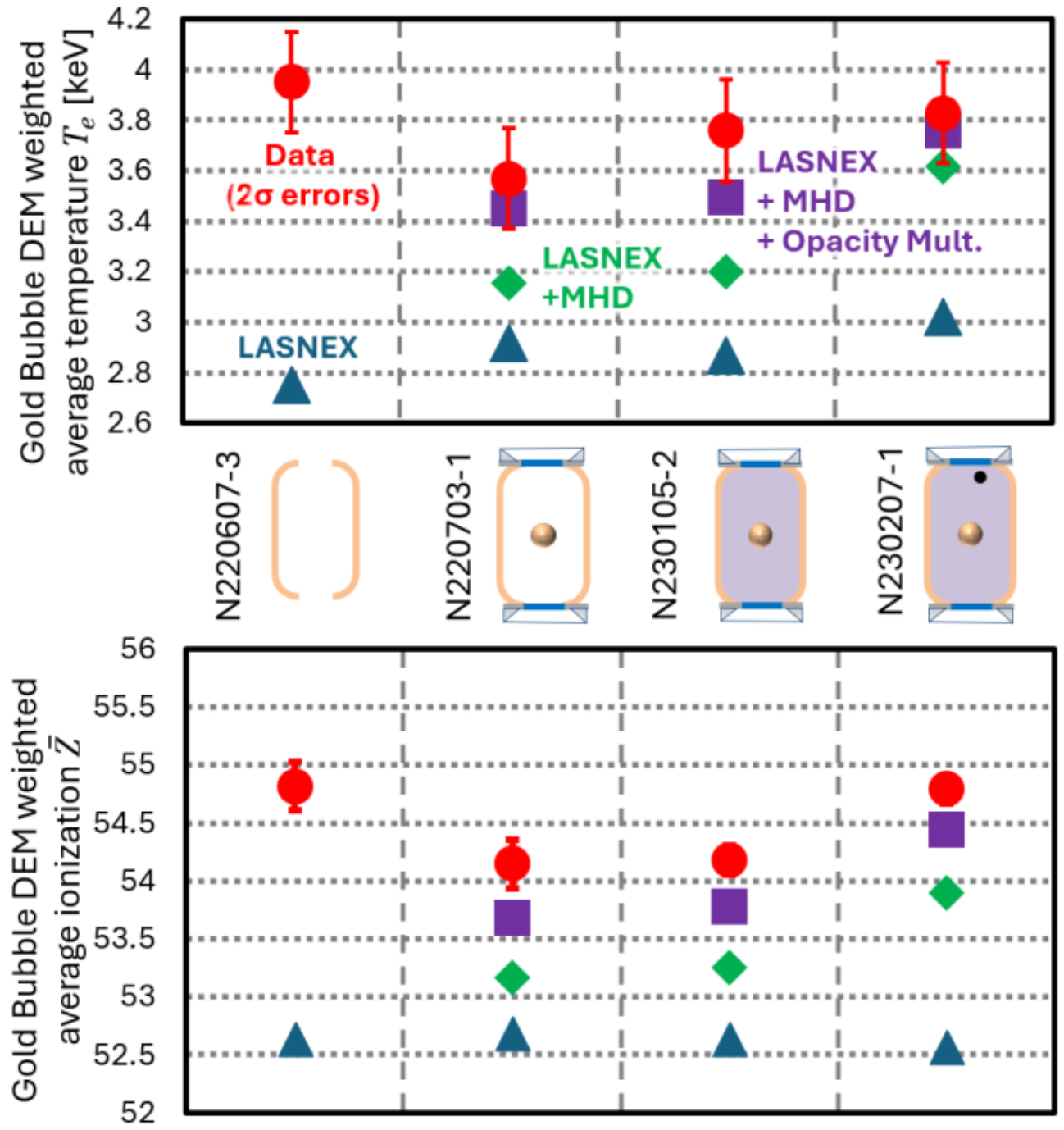

FIG. 11 $T_e$ and $\bar{Z}$ measurments of the gold bubble, resulting from analysis of NSS spectra captured in the BAH campaign. LASNEX LHT simulations systematically underestimate both the mean temperature and ionization state of the gold bubble plasma. Simulations that include MHD and the $\kappa_m$ opacity multiplier markedly improve agreement.

## VI. OPACITY MULTIPIERS AS A MEANS TO IMPROVE AGREEMENT BETWEEN EXPERIMENT AND SIMULATION

The data presented in sections V.B & V.D leads us to the conclusion that there are deficiencies in the NLTE radiation treatment in the current Lasnex Hohlraum Template (LHT) common model. These results corroborate recently published data from ViewFactor experiments, where it was found that an opacity-multiplier of $\kappa_M = 0.8$ could greatly improve the agreement between simulations and experiments[32]. This opacity multiplier is an empirical tuning factor that is used to reduce the plasma opacity for all photon energies > 1.8 keV. The intended effect of this multiplier is to reduce the emissivity of high temperature plasma in the gold bubble, effectively reducing the conversion efficiency from absorbed laser power to hohlraum x-ray power ($\eta_{laser}$). The reduced high energy x-ray emissivity will be accompanied by an increase in the gold bubble $T_e$, reflecting reduced radiative losses. For the BAH experiments using a 3.11 mm LEH we found that an opacity multiplier of $\kappa_M = 0.87$ brought the simulated M-band x-ray emission measured using Dante-1 into good agreement with our experimental measurements, while maintaining reasonable agreement with the measurements of total x-ray emission. Simulated drive parameters for simulations using the $\kappa_m = 0.87$ multiplier are included in FIG. 8 a), and the discrepancies for simulations using the multiplier are plotted in panel d). These can be compared to the discrepancies for the "vanilla" LHT simulations plotted in panel b). Overall, we see that the peak simulated $T_{R1}^S$ for Dante-1 decreased by ~ 5-10 eV for each experiment. This was accomplished through a ~30% decrease in the emission in the 2 – 4 keV energy band. Overall agreement for Dante-1 is much improved with the use of the multiplier. We note here that the Dante-2 measurements of the radiation temperature are actually in worse agreement with the simulations using the opacity multiplier potential reasons for this will be discussed later in this section.

Inspecting FIG. 11, we see that the default LHT simulations, plotted in green, consistently underpredicted gold bubble $T_e$ as measured by NSS. Simulated $T_e$ was systematically ~ 0.7 - 1 keV cooler than measured across a broad range of target designs. This difference in temperature was attended by a similar discrepancy in average ionization state. The simulations plotted in green used the Lasnex MHD package with a Nernst multiplier of 0.1. Adding this heat transport model with its severe restriction on the Nernst term is not enough to bring the simulated $T_e$ into agreement with the spectroscopic measurements. For the simulations plotted in purple, the Nernst multiplier was set to unity and an opacity multiplier $\kappa_m = 0.87$ was adopted. The introduction of this multiplier greatly improved the agreement between the simulated gold bubble plasma conditions and experiment. In conclusion, while adding MHD with a restricted Nernst term to the LHT simulations improved agreement significantly, the addition of the $\kappa_m$ multiplier improved agreement to the point where temperatures were matched within the measurement uncertainties without requiring any severe Nernst restrictions.

To assess the effect of the $\kappa_m$ multiplier on bang time we select the example of experiment N230207-1. This experiment is selected as it is the only experiment to use a full three-shock drive pulse. Other experiments in the campaign that measured a bang time used a shorter drive pulse and therefore had much longer implosion coast times than is typical of ID-ICF target designs. These experiments had very large bang-time discrepancies and it was unclear if the long coast time was a factor. The LHT simulation bang-time discrepancy for N230207-1 was 330+50-150 ps. For the simulation using the $\kappa_m = 0.87$ multiplier the bang time discrepancy was reduced to 100+50-150 ps, i.e. in agreement within the uncertainty of our measurement. It is clear that the reductions in high energy photon emission have the capacity to address a large portion of the historical bang-time discrepancy.

As in the ViewFactor experiments[32], there is some evidence that the best choice of $\kappa_m$ to match Dante measurements for a specific experiment may vary with the LEH size – we see some discrepancy in the M-band measurement for the single measurement we have with the 4.6mm dia. LEH. This is best seen by comparing the M-band discrepancies between experiments N240102-1 and N220214-2, as the only change between these experiments is the reduction in LEH size. The simple opacity multiplier as currently implemented reduces emission for radiation > 1.8 keV from all sources in the hohlraum, including thermal contributions from the dense wall plasma. The sensitivity to LEH size therefore likely reflects changes in the portion of the Dante line of sight that views the gold bubble.

These arguments may also be used to explain why the introduction of the opacity multiplier improves agreement for Dante-1 but degrades agreement for Dante-2. While Dante-1 has a clear view of the outer beam ring and gold bubble, Dante-2 mainly sees a portion of the hohlraum wall that is only indirectly heated (see FIG. 7). The opacity multiplier can therefore simultaneously improve Dante-1 agreement by reducing the M-band emission from the gold bubble, while degrading Dante-2 agreement by reducing the tail of the thermal emission.

Overall, it is clear that further improvements in NLTE opacity modeling will be required to develop a reliable and consistent emission model that accurately matches measured M-band emission for all targets and accurately predicts capsule bang-times. As things stand, the sensitivity of the required $\kappa_m$ to the specific diagnostic view of the target precludes its adoption as a fully predictive solution to hohlraum drive discrepancy problem. Despite this caveat, the introduction of an M-band multiplier appears to be capable of addressing a large part of the historical discrepancies in x-ray drive, gold bubble conditions and bang-time. This represents a great improvement in fidelity over the existing LHT simulation configuration and this is approach is therefore sure to prove useful in pushing forward the frontier of inertial confinement science.

## VII. CONCLUSIONS

The Build-A-Hohlraum campaign has provided invaluable insights into the underlying causes of historical bang-time discrepancies between simulation and experiment. Little variation in the discrepancies between simulated and experimentally measured x-ray emission were seen over the broad range of target complexity sampled in the campaign. Agreement between simulated and experimental x-ray drive was unaffected by the addition of the LEH hardware, gas fill and capsule. While the electron temperature in the gold bubble was observed to be higher than predicted by simulation, this enhanced temperature is seen for all target types, including vacuum hohlraums. The data is therefore inconsistent with the hypothesis that hohlraum energy losses due to mix effects are responsible for a significant portion of the bang-time discrepancy.

Our data is also inconsistent with the hypothesis that errors in the modeling of LEH closure are a key driving factor for the bang-time discrepancy. Across our suite of experiments, we see a consistent pattern, with simulations overpredicting the LEH size early in the drive, approximately matching LEH size at peak drive and slightly underpredicting size after the laser drive ends. These observations cannot explain the bang-time discrepancy. The overpredictions of the LEH size ought only to drive reduced energy coupling to the capsule in simulations.

Total x-ray emissivity was measured to be ~6 % lower than simulated across the campaign but this discrepancy can be entirely attributed to a 30 % overprediction of emission in the higher energy 2 – 4 keV band in simulations. Once this discrepancy is accounted for, we find that the simulated LTE blackbody x-ray emission is in good agreement with simulation. We therefore reject the hypothesis that errors in the modeling LTE hohlraum wall emission are responsible for a significant portion of the bang-time discrepancy.

The overpredicted 2 – 4 keV emission seen in our experiments is shown to be sufficient to explain most of the historical bang-time discrepancy. This emission discrepancy is accompanied by an under prediction of gold bubble electron temperatures and ionization states infered via spectroscopy. These observations are consistent with errors in the NLTE modeling of M-band emission from the hot gold bubble plasma. This result corroborates the conclusions of another recent paper[32], which reported on radiation drive measurements from ViewFactor targets.

Following the approach taken in the ViewFactor study, it was found that it was possible to greatly improve agreement between experiment and simulation through the introduction of an M-band opacity multiplier for x-rays > 1.8 keV, in our case $\kappa_m = 0.87$. Further, this allowed for the adoption of a local MHD heat transport model that largely matched inferred gold bubble plasma conditions within measurement errors. For simulations with a fueled capsule, introduction of this multiplier reduced the bang-time discrepancy from 300 ps to 100 ps, within our measurement uncertainty.

Unresolved discrepancies in the Dante-2 emission measurements may point to remaining deficiencies in the modeling of the LEH hardware. Our measurements have shown that the current model for the hardware is somewhat deficient.

Overall, the results of this campaign provide a clear indication of the areas in need of improvement. These results will be used to motivate both improvements to our NLTE radiation models and the design of new focused experiments to further investigate these discrepancies.

## ACKNOWLEDGEMENTS

This work was performed under the auspices of the U.S. Department of Energy by Lawrence Livermore National Laboratory under Contract DE-AC52-07NA27344 and by General Atomics under Contract 89233119CNA000063.